\documentclass[a4paper,11pt]{article}
\pdfoutput=1 % if your are submitting a pdflatex (i.e. if you have
\usepackage{jcappub} 
\usepackage[T1]{fontenc} 

\usepackage{natbib}
\usepackage{cancel}
\usepackage{enumitem}    
\usepackage{amsthm}         
\usepackage{amsmath} 	   
\usepackage{amssymb}       
\usepackage{amsfonts}
\usepackage{graphicx} 
\usepackage{dblfloatfix}	    
\usepackage{verbatim}
\usepackage{epstopdf}
\usepackage{tensor}
\usepackage{epsfig,multicol,bbm}
\usepackage{color}
\usepackage{colortbl}
\usepackage{float}
\usepackage{multirow}
\usepackage{array}

\usepackage{subfig}

\allowdisplaybreaks

\title{Dynamical attractors in contracting spacetimes dominated by kinetically coupled scalar fields}
\author[a,b,1]{Anna Ijjas,}
\note{Corresponding Author}
\author[c]{Frans Pretorius,}
\author[c]{Paul J. Steinhardt}
\author[d]{and David Garfinkle}

\affiliation[a]{Max Planck Institute for Gravitational Physics, Hanover, 30167, Germany}
\affiliation[b]{Institute for Theory and Computation, Harvard University, Cambridge, MA, 02138, USA}
\affiliation[c]{Department of Physics, Princeton University, Princeton, NJ, 08544, USA}
\affiliation[d]{Department of Physics, Oakland University, Rochester, MI 48309, USA}

\emailAdd{anna.ijjas@aei.mpg.de}

\abstract{We present non-perturbative numerical relativity simulations of slowly contracting spacetimes in which the scalar field driving slow contraction is coupled to a second scalar field through an exponential non-linear $\sigma$ model-type kinetic interaction.  These models are important  because they can generate a nearly scale-invariant spectrum of super-Hubble density fluctuations fully consistent with cosmic microwave background observations. 
We show that the non-linear evolution rapidly approaches a  homogeneous, isotropic and flat Friedmann-Robertson-Walker (FRW) geometry 
for a wide range of inhomogeneous and anisotropic initial conditions.  
Ultimately, we find, the kinetic coupling causes the evolution to deflect away from flat FRW and towards a novel Kasner-like stationary point, but in general this occurs on time scales that are too long to be observationally relevant. 
}


\begin{document}
\maketitle 
\raggedbottom

\section{Introduction}
\label{sec_intro}

Cosmic microwave background observations have shown that the spectrum of temperature anisotropies is nearly scale invariant and gaussian with an average amplitude of one part in hundred thousand and no detectable B-mode polarization  thus far. According to our current understanding, the temperature anisotropies are an imprint of primordial curvature fluctuations sourced by quantum excitations of one or more scalar fields, the energy density of which dominated the early-universe \cite{Bardeen:1983qw,Mukhanov:1988jd}. 

In order for the curvature fluctuations to match the observed spectrum, by the time the  modes are generated, the cosmological background must be smooth and flat as described by a Friedmann-Robertson-Walker (FRW) geometry with line element ${\rm d}s^2 = -{\rm d}\tau^2+a(\tau)^2{\rm d}x^i {\rm d}x_i$, where $a(\tau)$ is the FRW scale factor.  
Then, once the modes are generated, the smoothing mechanism must continue for at least 60 additional $e$-folds.
Neither of these two conditions is trivial to satisfy. For example, if the FRW solution is a robust attractor for a wide range of initial conditions, most of the volume is eventually smoothed and flattened by the time the smoothing phase ends; but, if the smoothing is {\it not sufficiently rapid}, most volume converges to a smooth and flat FRW spacetime with only a few $e$-foldings remaining before the end of smoothing. In this case, modes generated around the 60 $e$-fold mark would carry an imprint of an unsmoothed inhomogeneous and/or anisotropic geometry and would thus not match the observed spectrum. 

Slow contraction \cite{Cook:2020oaj}, a primordial phase that connects to the hot expanding phase through a gentle classical bounce, has been shown to be both robust and rapid.  The phase can be reached via a canonical scalar field $\phi$ that is minimally coupled to Einstein gravity and has a negative potential $V(\phi)$.  The scalar field energy density naturally evolves to dominate the total stress-energy while driving the geometry to a smooth and flat FRW space-time.  The robustness of slow contraction as a dynamical attractor -- its insensitivity to initial conditions including those that lie outside the perturbative regime of FRW spacetimes -- was recently shown in Refs.~\cite{Ijjas:2020dws,Ijjas:2021gkf,Ijjas:2021wml};  the remarkable rapidity, with smoothing and flattening typically occurring  within less than 10 $e$-folds of contraction of the Hubble radius, was demonstrated in Ref.~\cite{Ijjas:2021wml}.   

The goal of this paper is to examine whether the powerful smoothing property remains in cases in which the smoothing scalar field $\phi$ is coupled to a second scalar field $\chi$ through an exponential non-linear $\sigma$ model-type kinetic interaction.  These models are important because it has been shown that they can lead to the generation of a nearly scale-invariant spectrum of super-Hubble density fluctuations fully consistent with cosmic microwave background observations \cite{Li:2013hga,Levy:2015awa}.  More generally, this study is important for exploring whether slow contraction remains a powerful, robust and rapid dynamical attractor even when there is an exponentially strong kinetic interaction with a secondary scalar field and whether any distinctive features result compared to the case of scalar fields with canonical  kinetic energy density.

As demonstrated in Ref.~\cite{Ijjas:2021gkf},
the key to rapid and robust smoothing is {\it ultralocality}: Starting with arbitrary initial data that lies outside the perturbative regime of the FRW state, contracting spacetimes rapidly evolve to an anisotropic and spatially curved state where gradients, {\it i.e.}, spatial derivatives, are suppressed relative to the other so-called {\it velocity} contributions. These causally separated regions then each independently evolve to the homogeneous, isotropic and spatially flat FRW state which is the only stable stationary point of the ultralocal limit. 

Once the flat FRW state is reached, physical distances between objects shrink proportional to the scale factor 
$a(\tau) $
and exponentially slower  than the Hubble radius, 
 \begin{equation}
|H^{-1}| \equiv |{\rm d}{\rm ln}\,a(\tau)/{\rm d}\tau|^{-1} \propto  a^{\epsilon_{\phi}}, 
 \end{equation}
where $ \epsilon_{\phi} \gg3$; hence the name: slow contraction. 
 The rate at which $|H^{-1}|$ contracts is determined by the effective equation of state associated with the scalar field on an FRW background:
\begin{equation}
\epsilon_{\phi} \equiv \frac32\left(1+\frac{p_{\phi}}{\rho_{\phi}}\right) = 3\times \frac{\frac12 \phi'^2}{\frac12 \phi'^2 + V(\phi)},
\end{equation}
where $p_{\phi}=\frac12 \phi'^2 - V(\phi) $ is the co-moving `pressure' and $\rho_{\phi} = \frac12 \phi'^2 + V(\phi)$ the co-moving energy density of the scalar field $\phi$ in the homogeneous FRW limit and prime denotes differentiation with respect to $\tau$.

For a negative exponential potential, 
\begin{equation}
V(\phi) = -V_0 e^{-\phi/M}, \quad V_0>0,
\end{equation}
as will be considered throughout this paper,  the scaling attractor solution of the Einstein-scalar system in the flat FRW limit is given by
\begin{equation}
\label{FRW-bg}
a(\tau) = (-\tau)^{1/\epsilon_{\phi}},
\quad \phi(\tau) = M_{\rm Pl} \sqrt{\frac{2}{\epsilon_{\phi}}}\times {\rm ln} (-A\,\tau),
\quad  \epsilon_{\phi} = \frac12 \times \left(\frac{M_{\rm Pl}}{M}\right)^2,
\end{equation}
where $A=M_{\rm Pl}{}^{-1} \epsilon_{\phi}\sqrt{V_0/(\epsilon_{\phi}-3)} $ and $\tau$ is running from large negative to small negative values.
  (Here and throughout the paper, reduced Planck units, $M_{\rm Pl}{}^{-2} \equiv 8\pi G_{\rm N}$ with  $G_{\rm N}$ being Newton's constant, are used.)
For a potential with the characteristic mass scale $M \sim 0.1\,M_{\rm Pl}$, the effective equation of state $\epsilon_{\phi}=50$ is such that the scale factor contracts only by a factor of $2$ while the Hubble radius shrinks by a factor of $2^{50}$.

A particularly important feature of slow contraction is the fact that, because the Hubble radius shrinks much faster than the scale factor,  the wavelengths of fluctuations (which are proportional to $a(\tau)$)  necessarily end up on super-Hubble scales by the end of slow contraction. Unlike inflation, though, slow contraction is a `supersmoother,' meaning that adiabatic ({\it a.k.a.} curvature) modes decay, whether they are of classical or quantum origin  \cite{Creminelli:2004jg}.   The decay is the opposite of what occurs in expanding universes and is associated with the fact that the evolution of the adiabatic modes during slow contraction is subject to an  antifriction-like term ($H \ll 0$) that has the opposite sign than the friction-like term ($H\gg 0$) in the expanding case.   This is an important and appealing feature of slow contraction because it suppresses quantum runaway effects. %that can result in a multiverse.  
At the same time, a mechanism is needed to generate the spectrum of temperature anisotropies observed in the cosmic microwave background and the seeds for structure 
formation.  

In Refs.~\cite{Li:2013hga,Levy:2015awa}, it has been shown that entropy modes ({\it i.e.} pressure fluctuations on hypersurfaces of constant energy density) generated by quantum fluctuations of a second scalar field during slow contraction can fulfill this role.  First, like adiabatic modes, the entropic modes are generated by quantum fluctuations and their wavelengths also end up on super-Hubble scales.  
Second, unlike the adiabatic modes, the entropy modes  can experience a net red shift effect.  This can occur, for example, if  the modes are sourced by a light scalar field $\chi$ that is kinetically coupled to the background field $\phi$ through a non-linear $\sigma$-type interaction, {\it e.g.,}  %$\frac{1}{2}\kappa(\phi)\nabla_{\mu}\chi\nabla^{\mu}\chi$.   
\begin{equation}
\label{full-action}
{\cal S} = \int {\rm d}^4x \sqrt{-g}\Big({\textstyle \frac12}M_{\rm Pl}{}^2R - {\textstyle \frac12}  \nabla_{\mu}\phi\nabla^{\mu}\phi - V(\phi) - {\textstyle \frac12}  \kappa(\phi)\nabla_{\mu}\chi\nabla^{\mu}\chi - U(\chi)\Big),
\end{equation}
where $g$ is the four-metric determinant and $R$ the Ricci scalar.

Assuming an exponential coupling 
\begin{equation}
\kappa(\phi) = e^{-\phi/m} 
\end{equation}
with a characteristic mass scale $m\lesssim M$ as will be considered throughout this paper, the stable attractor solution of the Friedmann-scalar system of equations with $U(\chi)\simeq0$,
\begin{alignat}{1}
& 3M_{\rm Pl}^2H^2 =\frac12 \phi'^2 + \frac12 \kappa(\phi)\chi'^2 -V_0 e^{-\phi/M} ,\\
\label{phi-eq}
&\phi'' + 3H\phi' + \frac{V_0}{M}e^{-\phi/M} = \frac12 \kappa_{,\phi}\chi'^2,\\
\label{chi-eq}
&\chi'' + \left(3H + \frac{\kappa_{,\phi}}{\kappa}\phi' \right) \chi'=0,
\end{alignat} 
is that $\chi$ is constant ($\chi'\equiv 0$) while $a(\tau)$ and $\phi(\tau)$ evolve with time according to the scaling solution given by Eq.~\eqref{FRW-bg}.
The constant $\chi$ solution is stable \cite{Levy:2015awa} because, on an FRW background, the kinetic coupling $\frac12\kappa(\phi)\nabla_{\mu}\chi\nabla^{\mu}\chi$ enters the evolution equation~\eqref{chi-eq} of the $\chi$-field in a way that adds to the Hubble anti-friction  a friction-like term,
\begin{equation}
3H \rightarrow 3H + \frac{\kappa_{,\phi}}{\kappa}\phi' = \frac{1}{\epsilon_{\phi}(-\tau)}\left( \frac{M_{\rm Pl}^2}{m\times M} - 3\right) \gg 0.
\end{equation}
Since the combination is positive, the $\chi$ field's kinetic energy experiences a net damping as if it were in a de Sitter-like background and freezes out at some constant value. 
Notably,  
 quantum fluctuations of the $\chi$ field also experience a net de Sitter-like damping term $3H + \phi'/m \gg 0$ such that their amplitude grows, leading to a nearly scale-invariant and gaussian spectrum of entropy modes on super-Hubble wavelengths.  Finally, it has been shown that the entropy modes 
 can source curvature modes, {\it e.g.} when exiting slow contraction and entering the bounce stage, as shown in Refs.~\cite{Ijjas:2020cyh,Ijjas:2021ewd}.

The remaining question is whether adding a $\chi$ field with the exponential non-linear $\sigma$ model-type kinetic interaction with $\phi$
needed to generate a nearly scale-invariant and gaussian spectrum of curvature perturbations  preserves the robustness and rapidity of slow contraction as established for the single field scenario. 
In this paper, we address this issue by adapting the mathematical and numerical techniques developed for the single-field case in Refs.~\cite{Garfinkle:2008ei,Cook:2020oaj,Ijjas:2020dws,Ijjas:2021gkf,Ijjas:2021wml}. 

Our non-perturbative analysis yields some surprising results that could not be obtained using the conventional methods of cosmological perturbation theory.  First, in the special case where $\chi$ is precisely massless, we find that the evolution 
can be rapidly deflected away from the flat FRW stationary point if the characteristic scale $M$ associated with $V(\phi)$ is too close to the Planck scale ($M=1$), or equivalently, if $\varepsilon_{\phi}$ is not sufficiently large; instead, the evolution is driven towards 
a Kasner-like solution in which the 
gradient of  $\chi$,  $\bar{S}_{\chi}{}^x(\tau, x)$, is non-zero and time-independent  
\begin{equation}
\partial_{\tau}\bar{S}_{\chi}{}^x \equiv 0.
\end{equation}
%(where $S_{\chi} \equiv const.$),  
Second, in generic models which have moderately smaller values of $M \lesssim 0.1$ or weakly broken shift symmetry in $\chi$, {\it e.g.},  a small mass for $\chi$, we find that the deflection effect is strongly suppressed such that the evolution leads 
 to a long-lived state with negligible $S_{\chi}{}^x(\tau, x)$ and the FRW scaling solution in Eq.~\eqref{FRW-bg}, similar to the impressive single-field result.   This state persists long enough and has  just the conditions needed to generate a nearly scale-invariant spectrum of curvature perturbations that can account for temperature fluctuations observed in the cosmic microwave background. 
 %With this result, we demonstrate that, despite a non-trivial transfer of energy from the smoothing field to the secondary field through the kinetic interaction, the evolution generally converges to a smoothing slow contraction dynamical attractor.  
The kinetic coupling ultimately causes the evolution to deflect from flat FRW, but on a time scale that is too long to be relevant for  cosmologies that connect to the hot expanding phase through a smooth (non-singular) bounce.

\section{Numerical Scheme}
\label{sec_methods}

For the non-perturbative, numerical analysis, we shall adapt the  orthonormal tetrad form of the Einstein-scalar field equations corresponding to the action given in Eq.~\eqref{full-action},
\begin{alignat}{1}
\label{cov-eq1}
G_{\mu\nu} &= \nabla_{\mu}\phi\nabla_{\nu}\phi + \kappa(\phi)\nabla_{\mu}\chi\nabla_{\nu}\chi \\
&- \Big( {\textstyle \frac12} \nabla_{\sigma}\phi\nabla^{\sigma}\phi + {\textstyle \frac12}\kappa(\phi)\nabla_{\sigma}\chi\nabla^{\sigma}\chi + V(\phi) + U(\chi)\Big)g_{\mu\nu}
,\nonumber\\
\Box \phi &= V_{,\phi} + {\textstyle \frac12} \kappa_{,\phi}\nabla_{\sigma}\chi\nabla^{\sigma}\chi ,\\
\label{cov-eq3}
\Box \chi &= U_{,\chi} - \frac{\kappa_{,\phi}}{\kappa} \nabla^{\sigma}\phi \nabla_{\sigma}\chi,
\end{alignat}
 as developed for the single-field case. As per convention, $g_{\mu\nu}$ is the spacetime metric, and $G_{\mu\nu}$ is the Einstein tensor.  (Here and for the remainder of the paper, we express dimensional quantities in reduced Planck units where $M_{\rm Pl}=1$.)
 Throughout, spacetime indices $(0-3)$ are Greek and spatial indices $(1-3)$ are Latin. The beginning of the alphabet ($\alpha, \beta, \gamma$ or $a,b,c$) denotes tetrad indices and the middle of the alphabet ($\mu,\nu,\rho$ or $i,j,k$)  denotes coordinate indices.
 In the following, we only give a brief overview with the goal of making the paper self-contained. A comprehensive description of the formulation as well as a complete derivation of the evolution and constraint equations are provided in Refs.~\cite{Ijjas:2020dws,Ijjas:2021gkf}.

\subsection{Variables}

As with any tetrad formulation of the field equations, spacetime points are represented through a family  of unit basis four-vectors, or {\it vierbein}, $\{e_0, e_1, e_2, e_3\}$, where $e_0$ is the timelike four-vector and the spacelike four-vectors of the triad $\{e_1, e_2, e_3\}$ each lie in the rest three-space of $e_0$. The  local Lorentz frame set  by the tetrad basis is flat, {\it i.e.},
\begin{equation}
g_{\alpha\beta} = e_{\alpha}\cdot e_{\beta} = \eta_{\alpha\beta},
\end{equation}
with $\cdot$ denoting the inner product of the tetrad and $\eta_{\alpha\beta}={\rm diag}(-1,1,1,1)$ being the Minkowski metric. 

The {\it geometric variables} of the formulation are the sixteen tetrad vector components $\{e_{\alpha}{}^{\mu}\}$ and the twenty-four Ricci rotation coefficients
\begin{equation}
\gamma_{\alpha\beta\lambda} \equiv e_{\alpha} \nabla_{\lambda} e_{\beta},
\end{equation}
where $\nabla_{\lambda}$ is the projection of the spacetime covariant derivative $\nabla_{\mu}$ onto the tetrad $e_{\lambda}$, $\nabla_{\lambda} \equiv e_{\lambda}{}^{\mu}\nabla_{\mu}$;
\begin{equation}
K_{ab} \equiv - \gamma_{0ba}
\end{equation}
defines the nine components of the shear tensor; and
\begin{equation}
N_{ab} \equiv {\textstyle \frac12}\epsilon_b{}^{cd} \gamma_{cda},
\end{equation}
is the induced curvature tensor associated with the spatial triad where $\epsilon_{abc}$ denotes the Levi-Civita-symbol. The eighteen components of $K_{ab}$ and $N_{ab}$ are {\it dynamical} variables.
The three-vectors 
\begin{equation}
b_a \equiv \gamma_{a00}, \quad \Omega_a \equiv \frac12 \epsilon_a{}^{bc} \gamma_{cb0}
\end{equation}
are frame {\it gauge} quantities with $b_a$ defining the proper local acceleration of the congruence and $\Omega_a$ defining the local angular velocity of the spatial triad relative to Fermi-propagated axes.

The geometric variables must be supplemented by the variables describing the two scalars. These are the field distributions $\phi, \chi $, their velocities and gradients, respectively, as detailed below in Section~\ref{WS}.

\subsection{Gauge fixing}
\label{sec:gauge}

For our numerical scheme, we fix the six gauge degrees of freedom of the {\it tetrad frame} in a way that makes the connection of the geometric variables to physical quantities straightforward:
\begin{itemize}
\item[-] we fix the spatial triad $\{e_1, e_2, e_3\}$ to be inertially non-rotating a.k.a. {\it Fermi propagated} ($\Omega_a\equiv0$);
\item[-] we require the shear tensor to be symmetric ($K_{ab}\equiv K_{(ab)}$), meaning that the tetrad congruence is {\it hypersurface-orthogonal} such that it  defines a particular foliation of spacetime into spacelike hypersurfaces of constant time $\{\Sigma_t\}$ with $e_0$ being the future directed timelike unit normal to $\{\Sigma_t\}$ and the spatial triad vectors being tangent to $\{\Sigma_t\}$.
\end{itemize}
With this choice of tetrad frame gauge, $K_{ab}$ denotes the extrinsic curvature of $\{\Sigma_t\}$ and the components of $N_{ab}$ are the spatial (or intrinsic) curvature variables. 
Note that the acceleration vector $b_a$ is implicitly fixed by this gauge choice through
\begin{equation}
b_a \times e_0(x^0) = - e_ae_0(x^0),
\end{equation}
where $x^0$ is the time coordinate of $\{\Sigma_t\}$ and $\times$ denotes scalar multiplication.

Furthermore, we must write the tetrad evolution and constraint equations in the form of partial differential equations (PDEs) by way of which we shall numerically evolve the geometric and scalar field variables specified on an initial spacelike hypersurface. To do so is particularly straightforward given our choice of a hypersurface-orthogonal tetrad because, in this frame gauge, elements of the transformation matrix $\{\lambda_{\alpha}{}^{\mu}\}$ between tetrad and coordinate basis four-vectors, \begin{equation}
e_{\alpha}=\lambda_{\alpha}{}^{\mu}e_{\mu},
\end{equation}
are easily identified with quantities of the 3+1 Arnowitt-Deser-Misner (ADM) formalism, namely: 
\begin{equation}
\lambda_0{}^0=N^{-1},\quad \lambda_0{}^i = - N^i/N,\quad \lambda_a{}^0=0,\quad \lambda_a{}^i=E_a{}^i,
\end{equation}
where $N$ is the ADM lapse, $N^i$ the ADM shift and the coordinate metric is being given by $g^{\mu\nu}=\eta^{\alpha\beta}\lambda_{\alpha}{}^{\mu}\lambda_{\beta}{}^{\nu}$. Finally, the directional derivatives along the tetrads can be written as
\begin{equation}
D_0 = N^{-1}\big(\partial_t - N^i\partial_i \big), \quad D_a = E_a{}^i\partial_i.
\end{equation}

To fix the four {\it coordinate} gauge degrees of freedom, we specify the lapse function $N$ and the shift vector $N^i$ by requiring that
\begin{itemize}
\item[-] surfaces of constant time $\{\Sigma_t\}$ have constant mean curvature (CMC), {\it i.e.}, $\Theta^{-1}\equiv - \frac13 K_a{}^a = const$. The CMC slicing condition fixes the lapse function in that it leads to an elliptic equation~\eqref{Neqn-rs} for $N$; and
\item[-] the spatial coordinates are co-moving ($N_i=0$), {\it i.e.} constant along both the congruence and the foliation.
\end{itemize}

As emphasized previously in Refs.~\cite{Ijjas:2020dws,Ijjas:2021gkf}, the particular choice of our coordinate gauge has several advantages: since the trace of the extrinsic curvature $\Theta^{-1}$ is spatially uniform on each $\{\Sigma_t\}$, we can define the time coordinate $t$ to track $\Theta$, {\it i.e.},
\begin{equation}
e^t = {\textstyle \frac13} \Theta,
\label{timechoice}
\end{equation} 
such that, in the homogeneous limit, $\Theta$ is the Hubble radius $|H^{-1}|$. In addition, we can use $\Theta$ to rewrite our equations in terms of dimensionless Hubble-normalized variables , 
\begin{eqnarray}
N &\rightarrow& {\cal N}\equiv N/\Theta,\\
\{K_{ab}, N_{ab}, E_a{}^i\} &\rightarrow&  \{\bar{K}_{ab} , \bar{N}_{ab}, \bar{E}_a{}^i  \}
\equiv \{K_{ab}, N_{ab}, E_a{}^i\} \times \Theta  \,,
\\
\{V, U\} &\rightarrow& \{\bar{V}, \bar{U}\} \equiv \{ V, U\} \times\Theta^2, 
\end{eqnarray}
where ${\cal N}$ is the Hubble-normalized lapse and bar denotes normalization by the mean curvature $\Theta^{-1}$ on constant time hypersurfaces.  

The combination of the time coordinate $t$ tracking the three-curvature with Hubble-normalized variables  is particularly useful for our purposes to study the rapidity and robustness of slow contraction because it enables us to run the simulation for any finite period without encountering singular behavior or stiffness issues: first, the putative singularity is at $t\to-\infty$ and, second, the rapidly changing Hubble radius is not entering the numerical calculation as a dynamical variable, leaving the slowly changing scale factor as the only relevant dynamical factor.

\subsection{Evolution and constraint equations} \label{WS}

Taking everything together, the Einstein-scalar field evolution equations~(\ref{cov-eq1}-\ref{cov-eq3}) in Hubble-normalized, orthonormal tetrad form yield an elliptic-hyperbolic system:
The lapse ${\cal N}$ is defined at each time step through an elliptic equation,
\begin{equation}
\label{Neqn-rs}
- D^aD_a{\cal N} + 2 \bar{A}^b D_b {\cal N} + {\cal N} \Big(3 +
\bar{\Sigma}_{ab} \bar{\Sigma}^{ab} + \bar{W}_{\phi}^2  +\bar{W}_{\chi}^2 - {\bar V}(\phi)
-  \bar{U}(\chi) \Big) = 3 \,;
\end{equation}
while the evolution of the remaining geometric as well as the scalar field variables  is given by a hyperbolic system of PDEs:
\begin{alignat}{2}
%E_alpha^i
\label{E-eq}
&\partial _t \bar{E}_a{}^i &{}={}& \bar{E}_a{}^i - {\cal N} \left( \bar{E}_a{}^i + \bar{\Sigma}_a{}^b\bar{E}_b{}^i \right)
\,,\\
%A_alpha
&{\partial _t} \bar{A}_b &{}={}& \bar{A}_b + {\textstyle \frac12} \bar{\Sigma} _b{}^cD_c{\cal N} - D_b{\cal N} + {\cal N} \Big(
 {\textstyle \frac12} D_c \bar{\Sigma}_b{}^c - \bar{A}_b -  \bar{\Sigma}_b{}^c \bar{A}_c \Big)  
\,,\\
%N^alphabeta
& \partial _t \bar{n}^{ab} &{}={}& \bar{n}^{ab} 
- \epsilon^{cd ( a} \bar{\Sigma} _d{}^{b )} D_c {\cal N} + {\cal N} \Big( - \bar{n}^{ab}
+ 2 \bar{n}^{(a}{}_c \bar{\Sigma}^{b )c}
- \epsilon^{cd ( a} D_c \bar{\Sigma}_d{}^{b )} \Big)   
\,,\\
%Sigma_alphabeta
& \partial _t \bar{\Sigma}_{ab} &{} ={} & \bar{\Sigma}_{ab} 
+ D_{\langle a}D_{b\rangle} {\cal N} + \bar{A}_{\langle a} D_{b \rangle} {\cal N} 
+ \epsilon _{cd (a} \bar{n}_{b)}{}^{d}D^c{\cal N}
\,\\
& &{}-{}& {\cal N} \left( 3 \bar{\Sigma}_{ab} + D_{\langle a} \bar{A}_{b\rangle}
+ 2 \bar{n}_{\langle a}{}^c \bar{n}_{b\rangle c}
- \bar{n}_c{}^c \bar{n}_{\langle ab\rangle}
-  \epsilon _{cd (a} \Big( D^c \bar{n}_{b)}{}^d  - 2 \bar{A}^c \bar{n}_{b )}{}^d \Big) 
\right)
\nonumber\\
& &{}+{}& {\cal N} \Big( \bar{S}_{\phi}{}_{\langle a} \bar{S}_{\phi}{}_{b \rangle} + \bar{S}_{\chi}{}_{\langle a} \bar{S}_{\chi}{}_{b \rangle} \Big)
\nonumber
\,,\\
%\end{alignat}
%\begin{eqnarray}
&\partial _t \phi &{}={}& {\cal N} \bar{W}_{\phi} 
,\\
&{\partial _t} \bar{S}_{\phi}{}_a &{}={}& \bar{S}_{\phi}{}_a + \bar{W}_{\phi} D_a {\cal N}
+ {\cal N} \left( D_a \bar{W}_{\phi} - \left( \bar{S}_{\phi}{}_a + \bar{\Sigma}_a{}^b
\bar{S}_{\phi}{}_b \right) \right)
,\\
&{\partial _t} \bar{W}_{\phi} &{}={}& \bar{W}_{\phi} + \bar{S}_{\phi}{}^a D_a {\cal N} 
+ {\cal N} \left ( D^a \bar{S}_{\phi}{}_a - 3 \bar{W}_{\phi} - 2 \bar{A}^b \bar{S}_{\phi}{}_b 
- \bar{V}_{,\phi} 
\right )\\
& &{}+{}&  \frac12 {\cal N} \,\frac{\kappa_{,\phi}}{ \kappa} \Big( \bar{W}_{\chi}^2 - \bar{S}_{\chi}{}^a \bar{S}_{\chi}{}_a \Big )
,\nonumber\\
%chi-eq
\label{chi-eq}
&\partial _t \chi &{}={}& {\cal N} \,\frac{ \bar{W}_{\chi}}{\sqrt{\kappa(\phi)}}
,\\
%Schi-eq
\label{Schi-eq}
&{\partial _t} \bar{S}_{\chi}{}_a &{}={}& \bar{S}_{\chi}{}_a + \bar{W}_{\chi} D_a {\cal N}
+ {\cal N} \left( D_a \bar{W}_{\chi} - \left( \bar{S}_{\chi}{}_a + \bar{\Sigma}_a{}^b
\bar{S}_{\chi}{}_b \right) \right)
\\
& &{}+{}& \frac12 \frac { \kappa_{,\phi} }{\kappa} {\cal N} \Big(\bar{W}_{\phi}\bar{S}_{\chi}{}_a - \bar{W}_{\chi}\bar{S}_{\phi}{}_a \Big)
,\nonumber\\
\label{Wchi-eq}
 &   {\partial _t} \bar{W}_{\chi} &{}={}& \bar{W}_{\chi} + \bar{S}_{\chi}{}^aD_a {\cal N} 
+ {\cal N} 
\left ( D^a \bar{S}_{\chi}{}_a - 3 \bar{W}_{\chi} - 2 \bar{A}^b \bar{S}_{\chi}{}_b 
- \frac{\bar{U}_{,\chi}}{\sqrt{\kappa}}\right )
\\
& &{}+{}&   \frac12 {\cal N}\, \frac { \kappa_{,\phi} }{\kappa} \left( \bar{S}_{\phi}{}^a \bar{S}_{\chi}{}_a - \bar{W}_{\phi}\bar{W}_{\chi} \right) 
\nonumber,
\end{alignat}
where curved brackets denote symmetrization $X_{(ab)} \equiv {\textstyle \frac12}(X_{ab}+X_{ba})$ and angle brackets denote traceless symmetrization defined as $X_{\langle ab \rangle} \equiv X_{(ab)} - {\textstyle \frac13}X_c{}^c\delta_{ab}$. The geometric variables \begin{equation}
\bar{n}_{ab} \equiv \bar{N}_{(ab)}, \quad \bar{A}_b \equiv {\textstyle \frac12}\epsilon_b{}^{cd} \bar{N}_{cd}, 
\end{equation} 
are the symmetric and antisymmetric components, respectively of the Hubble-normalized, spatial curvature tensor $\bar{N}_{ab}$; $\bar{\Sigma}_{ab}$ is the trace-free extrinsic curvature tensor,
\begin{equation}
\bar{\Sigma}_{ab} \equiv \bar{K}_{ab} - 1.
\end{equation}
The variables $\bar{W}_{\phi}, \bar{W}_{\chi}$ denote the Hubble-normalized scalar field velocities.

In addition, the evolution equations are supplemented by a set of constraints which we shall use to specify the initial data as well as to verify convergence of the numerical computation:
\begin{alignat}{1}
\label{constraintG-rs}
&\bar{A}^b \bar{A}_b - {\textstyle \frac23 } D_b \bar{A}^b
+ {\textstyle \frac16 }  \bar{n}^{ab} \bar{n}_{ab} - {\textstyle \frac{1}{12} }  ( \bar{n}_a{}^a)^2 
+ {\textstyle \frac16 } \bar{\Sigma}^{ab} \bar{\Sigma}_{ab}  
\\
+& {\textstyle \frac16 } \left( \bar{W}_{\phi}^2+ \bar{S}_{\phi}{}^a\bar{S}_{\phi}{}_a \right)
+ {\textstyle \frac13} {\bar V}(\phi) 
+ {\textstyle \frac16 } \left(  \bar{W}_{\chi}^2 + \bar{S}_{\chi}{}^a\bar{S}_{\chi}{}_a\right)
+ {\textstyle \frac13} \bar{U}(\chi)
= 1
\nonumber
,\\
\label{constraintC-rs}
& D_b \bar{\Sigma}_a{}^b
 - 3 \bar{\Sigma}_a{}^b \bar{A}_b - \epsilon _{abc} \bar{n}^{bd} \bar{\Sigma}_d{}^c - \bar{W}_{\phi} \bar{S}_{\phi}{}_a -  \bar{W}_{\chi} \bar{S}_{\chi}{}_a
= 0
\,,\\
\label{constraintJ-rs}
& D_a \bar{n}^{ac} + \epsilon^{abc} D_a \bar{A}_b - 2 \bar{A}_a \bar{n}^{ac} = 0
\,,\\
\label{constraintCOM-rs}
& \epsilon^{abk} \left(D_a \bar{E}_b{}^l - \bar{A}_a \bar{E}_b{}^l \right) - \bar{n}^{kc}\bar{E}_c{}^l = 0
\,,\\
\label{constraintS-phi-rs}
& \bar{S}_{\phi}{}_a = D_a\phi  
\,,\\
\label{constraintS-chi-rs}
& \bar{S}_{\chi}{}_a =  \sqrt{\kappa(\phi)}\,D_a\chi 
\,.
\end{alignat}
The variables $\bar{S}_{\phi a}, \bar{S}_{\chi a}$ denote the Hubble-normalized scalar field gradients. Note that, for simplicity, we rescaled the velocity and gradient terms of the $\chi$ field with the non-linear $\sigma$-type kinetic interaction $\kappa(\phi)$, as can be seen, {\it e.g.}, in Eqs.~(\ref{chi-eq}) ~and~(\ref{constraintS-chi-rs}).

\section{Initial data}
\label{sec:ID}

To study the robustness of the kinetically-coupled two-field model as given through Eq.~\eqref{full-action} to cosmic initial conditions, it is essential to perform a large number of numerical relativity simulations corresponding to a wide range of initial conditions including those that lie far outside the perturbative regime of homogeneous and isotropic FRW spacetimes. 

As detailed in Refs.~\cite{Ijjas:2020dws,Ijjas:2021gkf}, our numerical scheme allows for the variation of all freely specifiable geometric and scalar field variables, $\{\bar{E}_a{}^i, \bar{n}_{ab}, \bar{A}_b, \bar{\Sigma}_{ab} \}$ and $\{ \phi, \chi, \bar{W}_{\phi}, \bar{W}_{\chi} \}$, respectively.   To ensure that the initial data satisfy  
the  constraint equations~(\ref{constraintG-rs}-\ref{constraintS-chi-rs}), in particular,  energy and momentum conservation, 
we adapt, as in our earlier work, the York method \cite{York:1971hw} that is commonly used in numerical relativity studies. (If the constraint equations are satisfied at the initial time, the Einstein equations propagate them such that they are satisfied at all later times, a condition that is checked numerically.)

Employing the same tetrad frame and coordinate gauge conditions that we detailed above in Sec.~\ref{sec:gauge}, we first fix the value of the inverse mean curvature $\Theta_0$ of the spatial hypersurface $\Sigma_{t_0}$ at some initial time $t_0$ and then define the three-metric of $\Sigma_{t_0}$ to be {\it conformally-flat}, {\it i.e.},
\begin{equation} \label{cf}
g_{ij}(t_0, {\bf x}) = \psi^4(t_0, {\bf x}) \delta_{ij},
\end{equation}
where the conformal factor $\psi$ is not a free function but is determined by an elliptic equation (given in Eq.~\ref{psi-eq} below) upon setting all other variables.  
Note that the conformally-flat metric choice  in Eq.~\eqref{cf} does not impose a limitation on cases that can be studied since it does not propagate; in fact, it is immediately violated after the first evolution step.

Our choice of $\Theta_0$ and $g_{ij}(t_0, {\bf x})$ fixes the coordinate components of the spatial tetrad basis vectors,
\begin{equation}
\label{init-E}
\bar{E}_a{}^i(t_0, {\bf x}) = \psi^{-2}(t_0, {\bf x}) \Theta_0 \delta _a{}^i,
\end{equation}
and the intrinsic curvature variables,
\begin{equation}
\label{init-n,a}
\bar{n}_{ab}(t_0, {\bf x}) = 0, \quad
\bar{A}_b(t_0, {\bf x}) = - 2 \psi^{-1}(t_0, {\bf x}) \bar{E}_b{}^i(t_0, {\bf x}) \partial _i \psi (t_0, {\bf x})
.
\end{equation}
Upon substitution of Eqs.~(\ref{init-E}-\ref{init-n,a}), it is straightforward to verify that the constraints~(\ref{constraintJ-rs} and \ref{constraintCOM-rs}) are trivially satisfied.

Furthermore, with a conformally-flat spatial metric, the momentum constraint~\eqref{constraintC-rs} reduces to the following simple relation,
\begin{equation}
\partial^b Z_{ab}^0 = Q_{\phi} \partial_a \phi + Q_{\chi} \sqrt{\kappa(\phi)}\partial_a \chi,
\label{divZ}
\end{equation}
where 
\begin{equation}
Z_{ab}^0 \equiv \psi^6(t_0, {\bf x})\bar{\Sigma}_{ab}(t_0, {\bf x}),
\end{equation}
 \begin{equation}
Q_{\phi}(t_0, {\bf x}) \equiv \psi^6(t_0, {\bf x})\bar{W}_{\phi}(t_0, {\bf x}),\quad 
Q_{\chi}(t_0, {\bf x}) \equiv \psi^6(t_0, {\bf x})\bar{W}_{\chi}(t_0, {\bf x}),
\end{equation}
denote the conformally rescaled Hubble-normalized shear and scalar field velocity variables, respectively. 

For simplicity and without loss of generality, we choose the initial value of $\phi$ and $\chi$ to be zero,
\begin{equation}
\phi(t_0, {\bf x}) = 0,\quad \chi(t_0, {\bf x}) = 0,
\end{equation}
turning the {\it momentum} constraint~\eqref{divZ} into the condition that the initial shear component $Z_{ab}^0$ be divergence free and giving us full freedom to choose the initial scalar field velocities $Q_{\phi}$ and $Q_{\chi}$.

We fix these quantities as follows:
\begin{equation}
\label{zic}
Z_{ab}^0 = \left(
\begin{array}{ccc}
{b_2} & \xi & 0\\
\xi & a_1\cos x + {b_1} & a_2\cos x\\
0 & a_2\cos x & -{b_1}-{b_2}-a_1\cos x
\end{array}\right),
\end{equation}
and 
\begin{alignat}{1}
\label{Qic}
Q_{\phi}(t_0, {\bf x}) &= {\Theta}_0 \Big (f_{\phi} \cos\big(\mu_{\phi} x + d_{\phi}\big) + Q_0\Big)
,
\\
\label{Qic2}
Q_{\chi}(t_0, {\bf x}) &= {\Theta}_0 \Big ( f_{\chi} \cos\big(\mu_{\chi} x + d_{\chi}\big) \Big).
\end{alignat}
%
%and the initial field value
%\begin{equation}
%\label{phiic}
%\phi(t_0, {\bf x}) = f_1 \cos(m_1 x + d_1).
%\end{equation}
%%
% IN THE CODE, a = a_1, and c = a_2;
% changed to this to avoid confusion with parameter c in
% the potential
%
where the parameters $\xi ,\, a_1, \, a_2,\  b_1, b_2, f_{\phi}, \mu_{\phi}, d_{\phi},  f_{\chi}, \mu_{\chi}, d_{\chi}$, and $ Q_0$  are constants.
%IN THE CODE: f_{\phi}=f_0, m_{\phi}=m_0, d_{\phi}=d_0
%
The sinusoidal form reflects the fact that, for the numerical simulation, we  choose periodic boundary conditions $0 \le x \le 2 \pi$ with $0$ and $2 \pi$ identified. 
For simplicity, all deviations from homogeneity are along a single spatial direction $x$, as in Ref.~\cite{Ijjas:2020dws}.

Finally, with the initial value of all freely specifiable geometric and scalar field variables $\Theta_0, \bar{E}_a{}^i, \bar{A}_b, \bar{n}_{ab}, Z_{ab}^0, \phi, \chi, Q_{\phi}, Q_{\chi}$ fixed and satisfying the constraint equations~(\ref{constraintC-rs}-\ref{constraintS-chi-rs}), we impose the remaining {\it Hamiltonian} constraint~\eqref{constraintG-rs} on these variables. This yields an elliptic equation for the conformal factor $\psi$,
\begin{equation}
\label{psi-eq}
\partial^a \partial_a \psi = {\textstyle \frac14 } \Theta_0^{-2} \left( 3  - \bar{V} - \bar{U} \right) \psi^5 
- {\textstyle \frac18 }  \left( \partial^a \phi \partial_a \phi \right) \psi
- {\textstyle \frac18 }\Theta_0^{-2}  \left( Q_{\phi}^2 + Q_{\chi}^2 + Z^{ab} Z_{ab} \right) \psi^{-7} ,
\end{equation} 
which we solve numerically.

\section{Numerical results}
\label{sec_results}

To numerically solve the Einstein-scalar field equations, we discretize the elliptic-hyperbolic system~(\ref{Neqn-rs}-\ref{Wchi-eq}) using second order accurate spatial derivatives and a three-step method for time integration employing the Iterated Crank-Nicolson algorithm. At each sub-step, we first solve the elliptic equation~\eqref{Neqn-rs} for the Hubble-normalized lapse ${\cal N}$ through a relaxation method and then update the hyperbolic equations~(\ref{E-eq}-\ref{Wchi-eq}) to the next Iterated Crank-Nicolson sub-step. In the simulations presented below, we use a grid of $4096$ points with $\Delta\, x = 2\pi /4096$ and  a Courant factor of $0.5$.
To demonstrate the convergence of our code, the error and convergence was analyzed for a broad range of examples using the same methods as detailed in the Appendices of Refs.~\cite{Ijjas:2020dws,Ijjas:2021wml}. To summarize those tests, our code shows no signs of numerical instability and exhibits clear second order
convergence at early times. At later times when a smooth, ultralocal spacetime develops, we empirically see the convergence improve to third order.

In this section, we present three  representative examples from our 
extensive  numerical studies of slow contraction with kinetically coupled scalar fields $\phi$ and $\chi$ given through the action in Eq.~\eqref{full-action}  with 
kinetic coupling 
\begin{equation}
\kappa(\phi) =   e^{-\phi/m}
\end{equation}
and potential energy densities 
\begin{equation}
V(\phi) = -V_0 e^{-\phi/M} \quad {\rm and}\quad 
U(\chi) = \frac12 m_{\chi} \chi^2, 
\end{equation}
respectively. 
We begin with the special case that $m_{\chi}=0$ and demonstrate a  subtle instability compared to the 
single-field case in Refs.~\cite{Ijjas:2020dws, Ijjas:2021gkf,Ijjas:2021wml}.  We then show that, for the generic 
case with  $m_{\chi}\ne0$, the instability is suppressed and
 robust and rapid convergence to a long-lived flat FRW state occurs. 
%These examples demonstrate a more subtle behavior compared to the 
%single-field case in Refs.~\cite{Ijjas:2020dws, Ijjas:2021gkf,Ijjas:2021wml}

%robustness Ijjas:2020dws
%mode and robustness Ijjas:2021wml
%ultralocal Ijjas:2021gkf
%supersmoothing

For each example, we show the evolution of the total scalar field energy density ($\Omega_{\phi-\chi}$), shear ($\Omega_s$) and curvature 
($\Omega_k$) 
contributions to the normalized energy density, defined as:
\begin{eqnarray}
\Omega_{\phi-\chi} &=& {\textstyle \frac16} \bar{W}_{\phi}^2 +  {\textstyle \frac16} \bar{S}_{\phi}{}^a \bar{S}_{\phi}{}_a +  {\textstyle \frac13} \bar{V}  
+ {\textstyle \frac16} \bar{W}_{\chi}^2 +  {\textstyle \frac16} \bar{S}_{\chi}{}^a  \bar{S}_{\chi}{}_a  +  {\textstyle \frac13} \bar{U} 
\\
\Omega_s &\equiv& {\textstyle \frac16} \bar{\Sigma}^{ab} \bar{\Sigma}_{ab}  
\\
\Omega_k &\equiv&
 - {\textstyle \frac23}D _b\bar{A}^b + \bar{A}^b\bar{A}_b 
 + {\textstyle \frac16}\bar{n}^{ab}\bar{n}_{ab} -  {\textstyle \frac{1}{12} } (\bar{n}_a{}^a)^2,
\end{eqnarray}
where $\Omega_{\phi-\chi} + \Omega_s + \Omega_k=1$.  We also formally define the separate contributions of the two 
fields,
\begin{eqnarray}
\Omega_{\phi}\equiv {\textstyle \frac16} \bar{W}_{\phi}^2 +  {\textstyle \frac16} \bar{S}_{\phi}{}^a \bar{S}_{\phi}{}_a
+  {\textstyle \frac13} \bar{V} ,
\\
\Omega_{\chi}\equiv {\textstyle \frac16} \bar{W}_{\chi}^2 + {\textstyle \frac16}  \bar{S}_{\chi}{}^a  \bar{S}_{\chi}{}_a  +  {\textstyle \frac13} \bar{U}.
\end{eqnarray}
Recall that the gradient $\bar{S}_{\chi}{}^a$ includes dependence on $\phi$ through the coupling $\kappa(\phi)$; see Eq.~\eqref{constraintS-chi-rs}.

The specify the parameters of the initial shear components $Z_{ab}^0$ and scalar field velocities $Q_{\phi}, Q_{\chi}$ given in Eqs.~\eqref{zic} and (\ref{Qic}-\ref{Qic2}), respectively, such that they correspond to initial conditions with highly non-perturbative deviations from a flat FRW spacetime:
\begin{alignat}{1}
&
\xi = 0.01,\quad a_1= 0, \quad a_2= 0.01, \quad  
b_1= -0.15, \quad b_2= 1.8,  
\\ %IN the CODE: \xi =kappa (only defined in init.f), a_1 = avec(2) and a_2 = avec(1) ,  b_1 = bvec(2) and b_2 = bvec(1) 
&f_{\phi}= 0.5, \quad \mu_{\phi}= 1, \quad d_{\phi}= -1.7, \quad Q_0 = 0.6,\\ 
&f_{\chi} = 0.1, \quad \mu_{\chi}= 1, \quad d_{\chi}= -1.7.
\end{alignat}
In addition, for the three representative cases, we fix the model parameters 
\begin{equation}
\label{boc}
V_0=0.1  \; {\rm and} \; M/m= 1.015, 
\end{equation}
where the ratio $M/m$ is chosen 
such that the predicted tilt of the temperature fluctuation spectrum matches current observations \cite{Planck:2018vyg}.  
Note that for $\xi ,a_1, a_2,b_1, b_2, f_{\phi}, \mu_{\phi}, d_{\phi}, Q_0$, and $V_0$, we chose the same values as in the single field case studied Ref.~\cite{Ijjas:2020dws}, to facilitate comparison.  

The time coordinate runs from initial time $t=0$ (or  $\Theta_0 = 3$)  towards $-\infty$ (or $\Theta \rightarrow 0$).  Equivalently, the time can be characterized by $n_H \equiv -t$, the number of $e$-folds of 
contraction of the inverse mean curvature $\Theta$, where $n_H$ runs from zero towards  $+\infty$.  In practice, models with a classical non-singular bounce undergo slow contraction until $n_H \approx 120$ before the bounce occurs \cite{Ijjas:2019pyf}.
For each example studied, the same initial data was evolved with several resolutions to confirm
second order convergence; the highest resolution has $4096$ points on the base level. 

\vspace{0.2in}
\noindent
The three representative cases correspond to different choices for $M$ and  $m_{\chi}$: 

\subsection{Case I: $M =0.2$ and $m_{\chi}=0$}
The choice $m_{\chi}=0$ is a {\it special case} where the action~\eqref{full-action} has a shift symmetry ($\chi \rightarrow \chi+$~const.).  The value of $M=0.2$ ($\epsilon_{\phi}=13$) was shown in the single field studies to be  just above the minimum required  for robust and rapid smoothing and flattening \cite{Ijjas:2020dws,Ijjas:2021wml}.  In this case with two fields and the special shift symmetry, though, we find a different outcome.

%%%%%
%%%
%%
%FIGURE1
\begin{figure*}[tb]
\begin{center}
\includegraphics[width=5.5in,angle=-0]{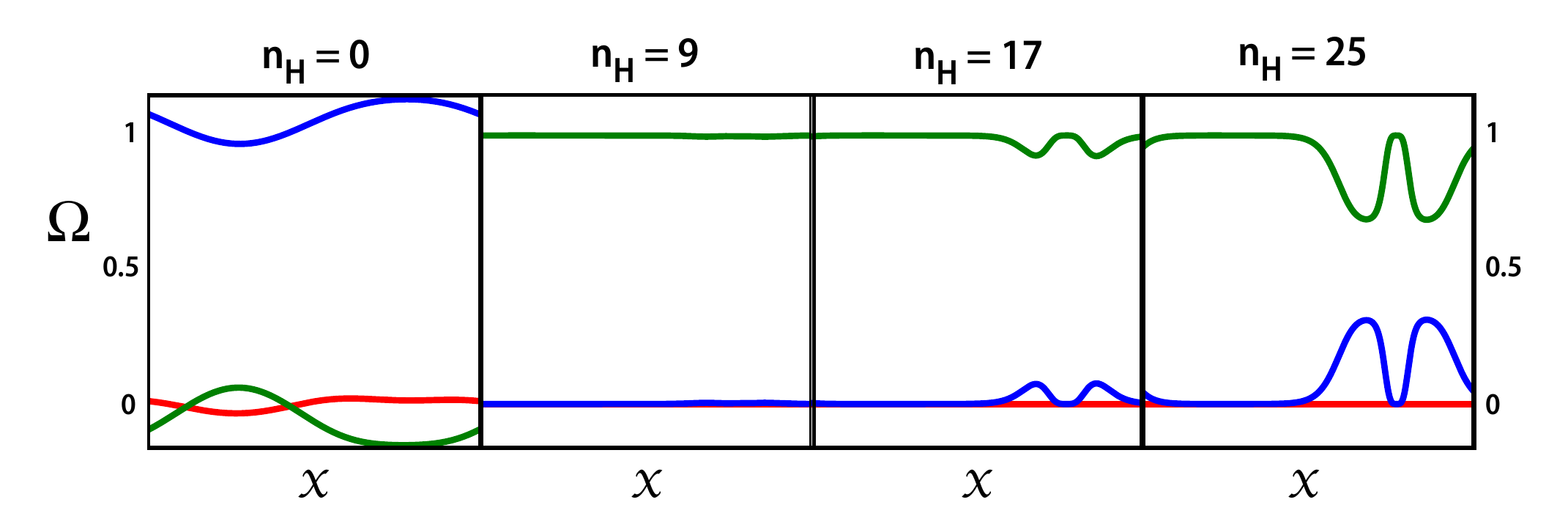}
\end{center}
\caption{For the case with $M=0.2$ and $m_{\chi}=0$, 
snapshots of the normalized energy density 
in  $\Omega_{\phi-\chi}$
(green), spatial curvature $\Omega_k$ (red) and
shear $\Omega_s$ (blue) 
for $0\le x \le 2 \pi$ at four different time steps $n_H= -t$, where $n_H$ is the number of $e$-folds of 
contraction of the inverse mean curvature $\Theta$.  
\label{Figure1}}
\end{figure*}

Fig.~\ref{Figure1} shows four snapshots of the evolution of the normalized scalar field energy density $\Omega_{\phi-\chi}$ (green), spatial curvature $\Omega_k$ (red) and
shear $\Omega_s$ (blue).  The first snapshot ($n_H=0$) shows the initial non-perturbative deviations from flat FRW.   By the second snapshot, a short time later ($n_H=9$), the spacetime approaches flat FRW with $\Omega_{\phi-\chi} \approx 1$ and $\Omega_s
\approx\Omega_k\approx 0$, seemingly similar to the single-field case.  But instead of remaining smooth and flat as found in the single-field case, something different occurs: the flat FRW phase is not stable, and the shear (blue curve) begins to grow, as illustrated in the last two snapshots.  At $n_H=25$, the geometry is spatially flat but  with a mix of field energy plus shear.  This combination was termed as `Kasner-like' in the single field case \cite{Ijjas:2020dws} where it was  found in certain examples with values $M$ bigger than $ 0.2\,M_{\rm Pl}$.  As we will see, though, something quite different is happening in the case of two kinetically coupled fields.  Note that spikes will form at locations where the $\chi$-gradient vanishes; {\it i.e.}, due to the ultralocal
nature of the evolution, such points get stuck at the FRW state, whereas adjacent regions,
destabilized by a non-zero $\chi$-gradient, transition to the final Kasner-like states. Thus, using
the terminology of Ref.~\cite{Ijjas:2020dws}, the state should be referred to as `Kasner-like (modulo spikes).' 

%FIGURE2
\begin{figure*}[t]
\begin{center}
\includegraphics[width=5.5in,angle=-0]{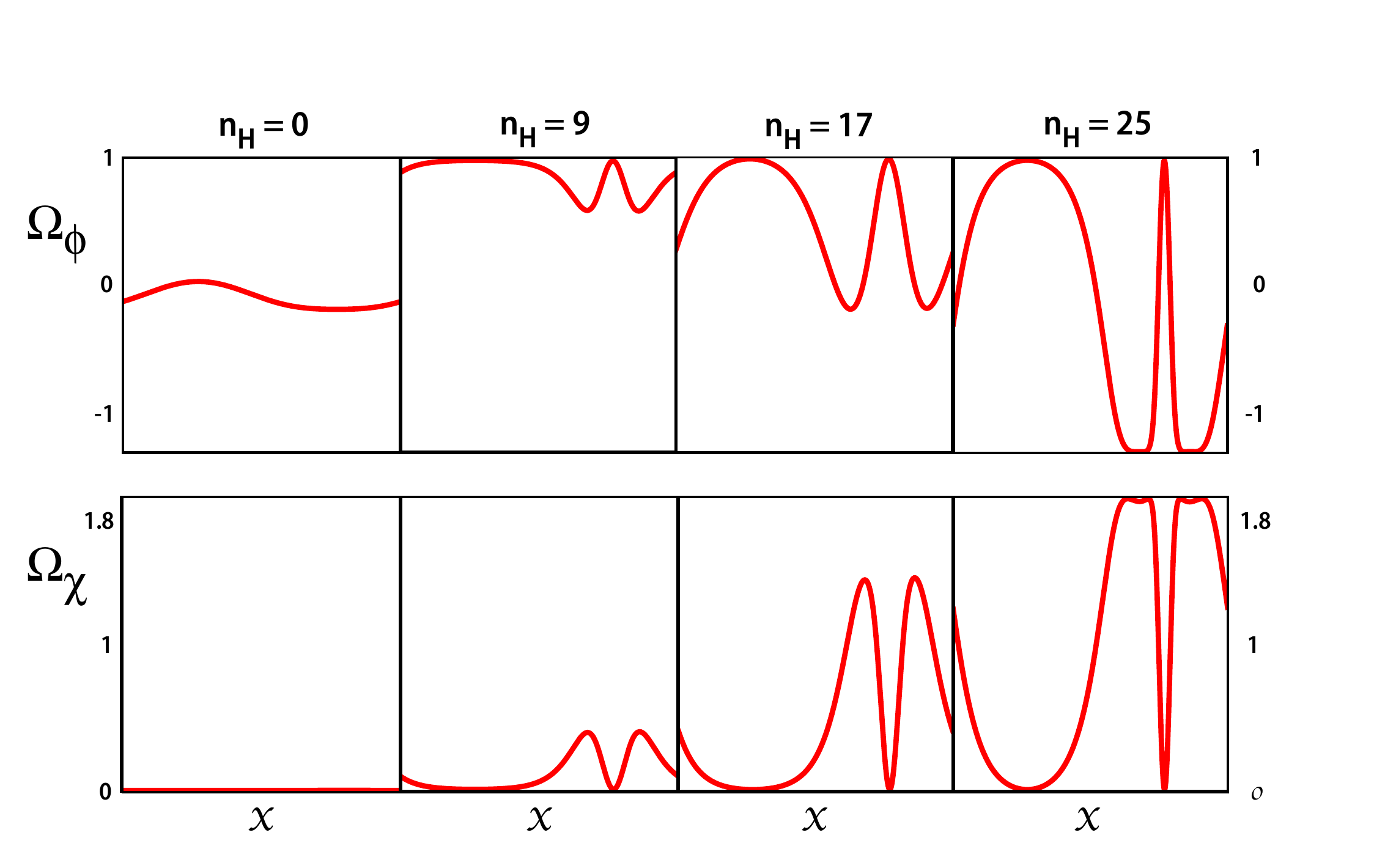}
\end{center}
\caption{For  the case with $M=0.2$ and $m_{\chi}=0$, 
snapshots of $\Omega_{\phi}$ (top row) and $\Omega_{\chi}$ (bottom row) 
at the same times as shown in Fig.~\ref{Figure1}.
\label{Figure2}}
\end{figure*}
Fig.~\ref{Figure2}, which tracks the evolution of $\Omega_{\phi}$ and $\Omega_{\chi}$ separately,  reveals more details. As the sequence progresses, it can be seen that $\Omega_{\phi}$ rapidly comes to dominate, but then energy is transferred to $\chi$ through their non-linear coupling such that $\Omega_{\chi}$ begins to grow.  More precisely, $\bar{W}_{\chi} \rightarrow 0$ almost immediately, by $n_H =2$, and remains negligible; however, the rescaled $\chi$-gradient $\bar{S}_{\chi}{}^{x} = \sqrt{\kappa(\phi)} D^x\chi$ grows rapidly proportional to the coupling factor, $\sqrt{\kappa(\phi)}$, that is growing rapidly due to the growth of $\phi$ which is quickly rolling downhill its potential $V(\phi)$.  (N.B. This growth of the gradient is real, {\it i.e.} not a frame effect, as explained in the Appendix~\ref{app:frame}.)
By the last snapshot $n_H=25$,  which is far short of the bounce, $\Omega_{\chi}$ grows to dominate over $\Omega_{\phi}$; and, then, looking back to the last snapshot  in Figure~\ref{Figure1}, we see that, at the same time, the gradient sources a growing shear component $\Omega_s$.  We stop the code at $n_H=25$ for reasons described in the next section where we also use analytics to determine how the evolution continues. 

\subsection{Case II: $M=0.1$ and $m_{\chi}=0$}  
Here again we consider the {\it special case} with $m_{\chi}=0$ in which the action~\eqref{full-action} has a shift symmetry. This example illustrates that
slightly decreasing $M$ (or, equivalently, increasing the equation of state during slow contraction $\epsilon_{\phi}$) significantly delays the onset of the instability but does not eliminate it, as illustrated by this example.
%
%FIGURE3
\begin{figure*}[tb]
\begin{center}
\includegraphics[width=5.5in,angle=-0]{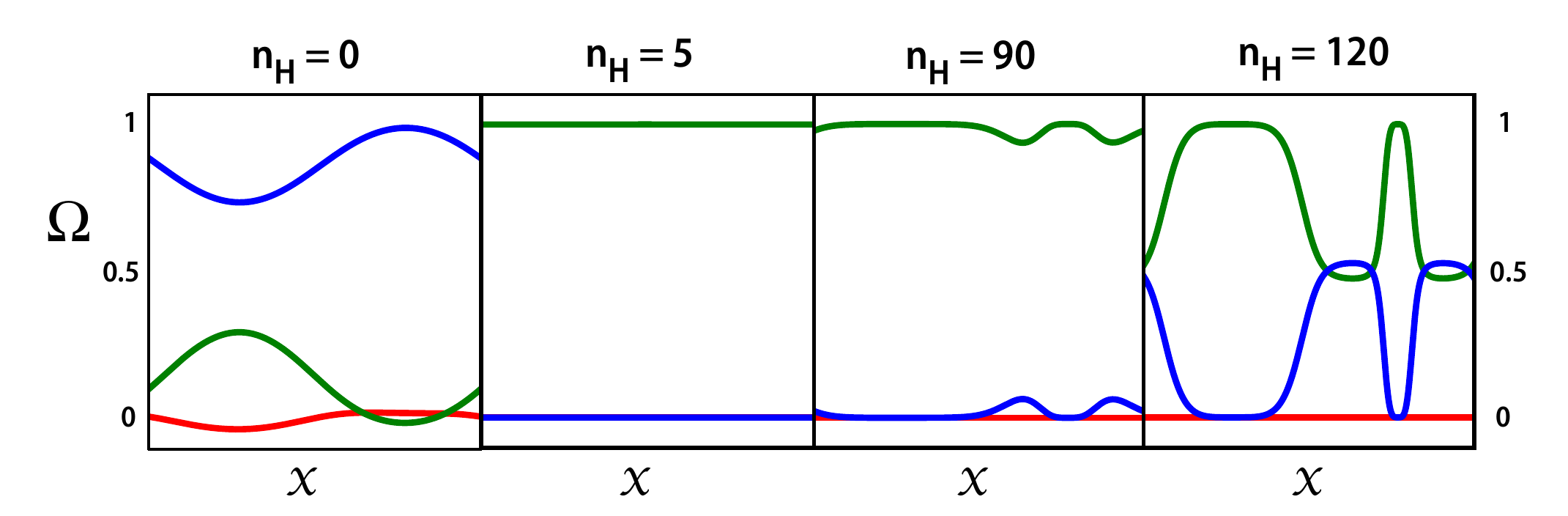}
\end{center}
\caption{For  the case with $M=0.1$ and $m_{\chi}=0$, 
snapshots of the normalized energy densities following the same color code 
as in Fig.~\ref{Figure1}.  Comparing to Fig.~\ref{Figure1}, one observes that 
decreasing $M$, or equivalently, increasing the equation of state $\epsilon_{\phi}$ of the $\phi$ field in Eq.~\ref{FRW-bg}, rapidly and robustly drives the universe towards FRW, but a non-zero $\chi$-gradient eventually destabilizes the FRW state. 
\label{Figure3}}
\end{figure*}
Fig.~\ref{Figure3} shows that the universe is rapidly flattened as in Case I, but in this case the shear $\Omega_s$ only begins to dominate for some ranges of $x$ at $n_H=120$.  If there were no bounce and the simulation were running for a yet longer range of $n_H$, the behavior would be similar to the Case I but shifted in time.   
As a practical matter, though, by decreasing $M$ modestly further, $M<1/15$, the instability could be delayed beyond $n_H=120$, which is more than sufficient for bouncing cosmologies with a smooth (non-singular) bounce (where the bounce occurs at $n_H <120$).

\subsection{Case III: $M =0.1$ and $m_{\chi}=300 \,\Theta_0^{-1}$}  
\begin{figure*}[tb]
\begin{center}
\includegraphics[width=5.5in,angle=-0]{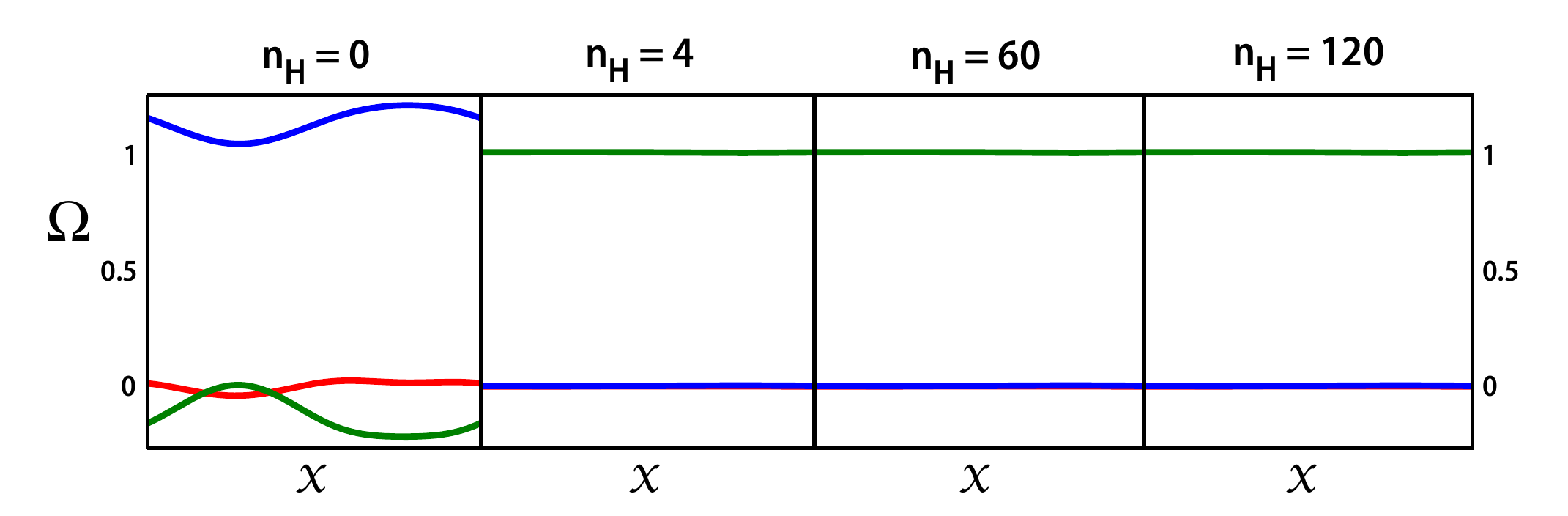}
\end{center}
\caption{For  the case with $M=0.1$ and $m_{\chi}=300\Theta_0^{-1}$, 
snapshots of the normalized energy density following the same color code 
as in Figs.~\ref{Figure1} and~\ref{Figure3}.  Compared to the case in Fig.~\ref{Figure3}, one observed that introducing a small $m_{\chi}$ suffices
to obtain robust and rapid smoothing.  
\label{Figure4}}
\end{figure*}

\begin{figure*}[t]
\begin{center}
\includegraphics[width=5.35in,angle=-0]{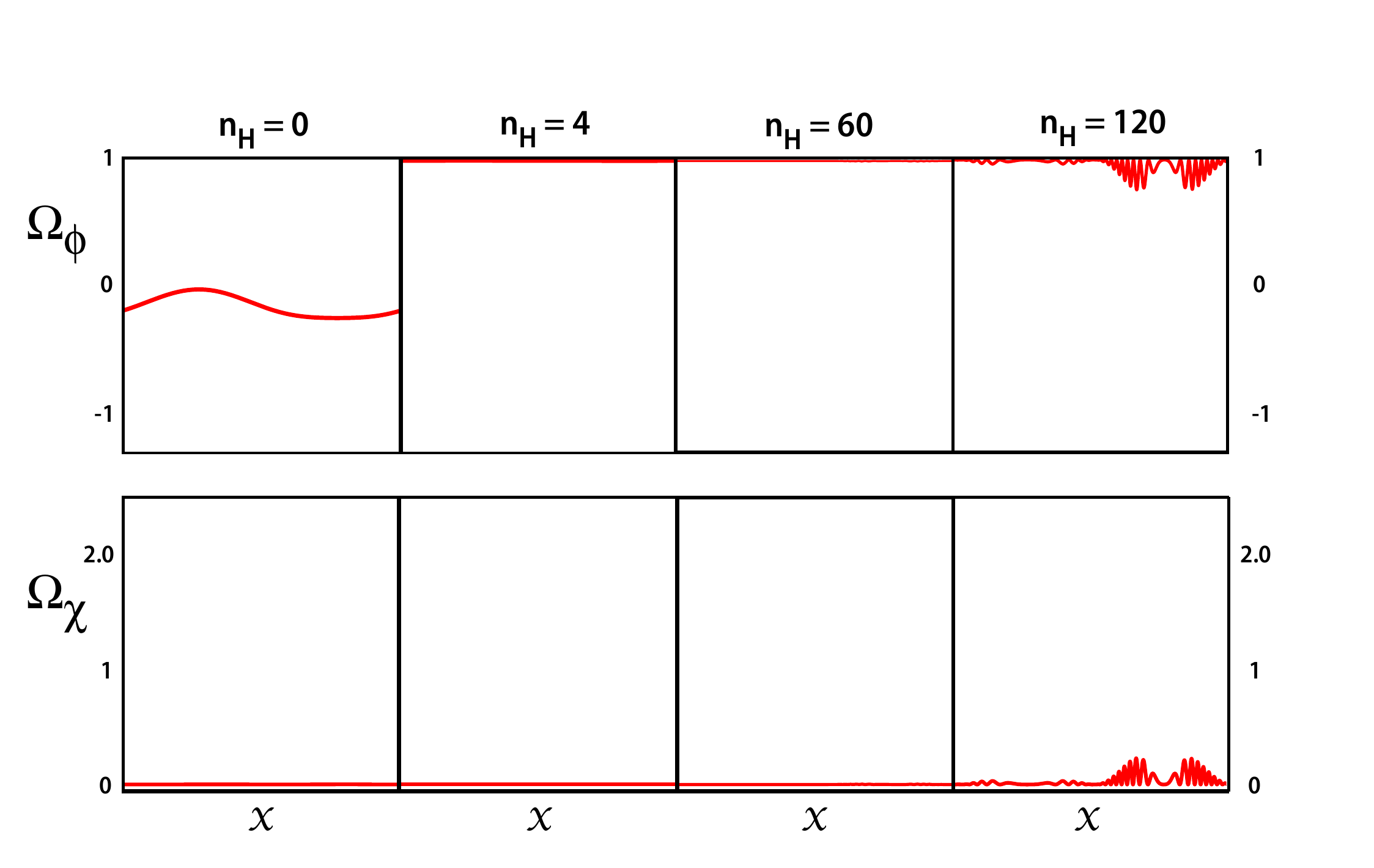}
\end{center}
\caption{For  the case with $M=0.1$ and $m_{\chi}=300$, 
 snapshots of $\Omega_{\phi}$ (top row) and $\Omega_{\chi}$ (bottom row) 
at the same times as in Fig.~\ref{Figure4}. 
\label{Figure5}}
\end{figure*}

Now we turn to the {\it generic} case of $m_{\chi} \ne 0$, thus breaking the shift symmetry, which we show also acts to suppress the instability.  (This case is generic because there is no reason to expect an exact shift symmetry in $\chi$ since there is no shift symmetry in $\phi$.)
Bearing in mind that the initial value of the inverse extrinsic curvature 
is $\Theta_0^{-1} \lesssim 10^{-42}$~GeV in bouncing cosmologies,
Figs.~\ref{Figure4} and~\ref{Figure5} show that it suffices to break the shift symmetry in $\chi$ with even a small mass $m_{\chi}$ in order to obtain a qualitatively different result.  
The reason will be explained in Section~\ref{sec:analytic} below.
%I could have inserted a statement here about $m_{\phi}$ only being smaller than $ \Theta^{-1}$ for $n_H<6$ in this case; which is enough to drive the universe to the flat FRW basin of attraction.  We had this on my list to include.  But I think this better belongs in your analytics discussion.  
% Since there is no shift symmetry in $\phi$ and no reason to expect an exact shift symmetry in $\chi$ alone, this example should be viewed as {\it generic} for slow contraction with two kinetically coupled fields. 

%FIGURE4

%%
%FIGURE5

Fig.~\ref{Figure4} shows that, beginning from the same initial conditions as in Cases~I and~II, the geometry robustly and rapidly (by $n_H=4$) converges to flat FRW, just as in the single field case.  (A similar result is found for $M=0.2$ for a slightly greater value of $m_{\phi}$.)  In particular, the gradients $\bar{S}_{\chi}^x$ and the shear $\Omega_s$ remain small over the $n_H=120$ e-folds of the simulation, sufficient for practical applications to bouncing cosmologies with a gentle (non-singular) bounce. 

Fig.~\ref{Figure5} shows a subtle difference from the single-field case, though.  Although the total
normalized scalar field energy density $\Omega_{\phi-\chi}$ is close to unity and the shear is negligible, the kinetic coupling between the two fields through the gradient $\bar{S}_{\chi}^x$  result in an exchange of energy between the two components, as shown by the compensating oscillations at $n_H=120$, albeit an exchange that generates negligible shear. These can be viewed as classically generated entropic fluctuations; in this example and for a wide range of $M$ and $m_{\chi}$, these fluctuations have an amplitude that  is irrelevant  compared to the quantum-generated entropic fluctuations on the length scales measured by cosmic microwave background observations.  Their presence is a sign that the evolution is beginning to deviate away from flat FRW; in principle, given additional time, the system would evolve to a Kasner-like fixed point as in the cases above (but this has no practical relevance for  cosmologies involving a slow contraction phase that connect to the hot expanding phase through a smooth non-singular bounce.)

\section{Analytic approximation}
\label{sec:analytic}
 
The numerical studies in the previous section show that slow contraction in the two-field models considered in this paper is in general a rapid and robust smoother over the time scales and length scales of interest for bouncing cosmologies.  
However, the three numerical results also show that there is a subtlety  that does not arise in the case of a single canonical scalar field with a steep negative potential.  Namely, a flat FRW state is not the ultimate fixed point attractor;  although the evolution initially approaches a flat FRW fixed point, it is subsequently deflected towards a homogeneous but anisotropic `Kasner-like' state.  

To understand this novel phenomenon, we study different stages in the evolution using analytic perturbative analyses.   These explain how the deflection from flat FRW arises and show that that the characteristic time scale for the deflection is in general  of ${\cal O}(100)$ or more $e$-foldings of contraction of the inverse mean curvature, which is of no practical relevance in cosmologies where slow contraction connects to the hot expansion phase through a smooth classical (non-singular) bounce.

\vspace{.1in}
\noindent
The evolution beginning from highly non-perturbative deviations from flat FRW involves four stages:  

\vspace{.1in}
\noindent
{\it Stage 1:}   During the first stage, the system evolves from the freely specified initial state (as detailed above in Sec.~\ref{sec:ID}) to one that is {\it ultralocal}, 
 meaning that terms involving spatial derivatives (gradients) are quickly dominated by velocity terms ({\it i.e.}, terms that do not involve spatial derivatives) as the evolution proceeds.  In particular, within only a few $e$-folds of contraction of the inverse mean curvature $\Theta$,
 \begin{equation}
\bar{E}_a{}^i \to 0, \; \bar{A}_b \to 0,\; D_a \phi \to 0,\; D_a \chi \to 0.
\end{equation}
Note that ultralocal does not mean flat FRW, as  extensively detailed in Ref.~\cite{Ijjas:2021gkf}.

\vspace{.1in}
\noindent
{\it Stage 2:}
For  $M\leq 0.2$ and $Q_0>0$, the same range considered in Ref.~\cite{Ijjas:2020dws}, the Einstein-scalar system~(\ref{E-eq}-\ref{Wchi-eq}) rapidly and non-linearly evolves from an ultralocal (but not FRW) state towards the flat FRW stationary point with 
\begin{equation}
 \label{FRW-sp-chi}
\bar{W}_{\chi} \simeq 0, \bar{S}_{\chi}{}^a\simeq0, \bar{U}_{,\chi}\simeq0,
\end{equation}
 \begin{equation}
 \label{FRW-sp}
 \bar{n}_{ab}\simeq0, \; \bar{\Sigma}_{ab}\simeq0, \; \bar{W}_{\phi}\simeq M^{-1}, \; \bar{V}\simeq3-\frac12 M^{-2},\; {\cal N}\simeq 2 M^2,
 \end{equation}
 just as we observe by $n_H=9$  in all of our simulations as illustrated in the second panels of
  Figures~\ref{Figure1}, \ref{Figure3} and~\ref{Figure4}.  
 
 \vspace{.1in}
\noindent
 The first two stages are very rapid (complete by $n_H \lesssim 10$ ) and are determined by the slow contraction sourced by the $\phi$-field.   As result, the evolution during these two stages is  similar to what is found for the single-field case. 
  
\vspace{.1in}
\noindent
{\it Stage 3:}  For the case of a single canonical scalar, flat FRW is a stable fixed point of the evolution. 
For the kinetically-coupled two-field models considered here, though, flat FRW is {\it not} a  stable fixed point.  Rather, there remain small deviations from flat FRW that eventually grow large enough to deflect the evolution from the flat FRW fixed point.   This effect can be  seen by perturbing the Einstein-scalar system of equations~(\ref{E-eq}-\ref{Wchi-eq}) around the flat FRW fixed point given by~Eqs.~(\ref{FRW-sp-chi}-\ref{FRW-sp}). 

Inspecting the linearized system,
\begin{alignat}{2}
%E_alpha^i
\label{E-eq-L}
&\partial _t \delta \bar{E}_a{}^i &{}={}& \Big(1- {\cal N}  \Big) \delta \bar{E}_a{}^i ,\\
%A_alpha
&{\partial _t} \delta\bar{A}_b &{}={}& \Big(1- {\cal N} \Big) \delta\bar{A}_b ,\\
%N^alphabeta
& \partial _t \delta \bar{n}^{ab} &{}={}& \Big(1- {\cal N} \Big) \delta \bar{n}^{ab} ,\\
%Sigma_alphabeta
& \partial _t \delta\bar{\Sigma}_{ab} &{} ={} &\Big(1- 3{\cal N}  \Big) \delta \bar{\Sigma}_{ab} 
,\\
%Wphi-eq
&{\partial _t} \delta \bar{W}_{\phi} &{}={}& \Big(1- 3{\cal N}  \Big) \delta \bar{W}_{\phi} 
- \Big(3\bar{W}_{\phi} - M^{-1} \bar{V}\Big) \delta {\cal N} + M^{-1} {\cal N}  \delta \bar{V}
,\\
%Sphi-eq
&{\partial _t} \delta \bar{S}_{\phi}{}_x &{}={}& \Big(1- {\cal N} \Big) \delta\bar{S}_{\phi}{}_x 
,\\
%chi-eq
\label{L-chi-eq}
&\partial _t \delta \chi &{}={}& {\cal N} \,\frac{ \delta \bar{W}_{\chi}}{\sqrt{\kappa(\phi)}}
\\
%Wchi-eq
\label{L-Wchi-eq}
 &   {\partial _t} \delta \bar{W}_{\chi} &{}={}& \left(1- {\cal N}\left(3 - \frac12 m^{-1}  \bar{W}_{\phi}\right)  \right) \delta \bar{W}_{\chi} 
- {\cal N} \frac{ \bar{U}_{,\chi\chi}}{\sqrt{\kappa}}\delta\chi
,\\
%Schi-eq
\label{L-Schi-eq}
&{\partial _t} \delta \bar{S}_{\chi}{}_x &{}={}&\left(1- {\cal N} \left(1 + \frac12 m^{-1}    \bar{W}_{\phi} \right)\right) \delta \bar{S}_{\chi}{}_x 
,
\end{alignat}
where $\delta$ denotes linear perturbations around the background solutions,   
it is immediately apparent that the perturbations of all geometric variables as well as the linearized scalar field variables $\delta \bar{W}_{\phi}, \delta \bar{S}_{\phi}{}_a$ form a closed system and decay at the same rate as in the single field case as $|t|$ grows, {\it i.e.},
\begin{equation}
\delta \bar{E}_a{}^i, \delta\bar{A}_b,  \delta \bar{n}^{ab}, \delta \bar{S}_{\phi}{}_x
\propto e^{(1-2M^2)\,t} ,
\end{equation}
\begin{equation}
\delta \bar{\Sigma}^{ab}, \delta \bar{W}_{\phi} \propto e^{(1-6M^2)\,t},
\end{equation}
where ${\cal N} = 2M^2\leq 2/5.2^2$; see Eq.~(\ref{FRW-sp}).   Hence the exponents $1-{\cal N} \geq 0.93, \;  1-3{\cal N}\geq 0.78$ are both positive definite, in agreement with the results found in Ref.~\cite{Ijjas:2020dws}. (Positive exponents correspond to decay because the time variable runs from $0$ towards $t \rightarrow -\infty$.)  
%Note that we used the lapse equation and the Hamiltonian constraint to eliminate the non-dynamical variables $\delta{\cal N}$ and $\delta\bar{V}(\phi)$.  

Solving the linearized equations~(\ref{L-chi-eq}-\ref{L-Wchi-eq}) around the flat FRW stationary point~\eqref{FRW-sp}, it is straightforward to verify that the kinetic interaction $\kappa(\phi)=e^{-\phi/m}$ makes the $\bar{W}_{\chi}=0$ solution stable. In matrix form, the closed system can be written as
\begin{equation}
\label{sta-eq-Wchi-sp}
\partial_t \begin{pmatrix}
\delta \chi\\[1em]
\delta \bar{W}_{\chi}
\end{pmatrix}
=
\begin{pmatrix}
0 & & 2M^2/\sqrt{\kappa(\phi)}\\[1em]
-2 M^2\bar{m}_{\chi}^2/\sqrt{\kappa(\phi)} & {\quad} & 1 - 6M^2 +  M/m
\end{pmatrix}
\begin{pmatrix}
\delta \chi\\[1em]
\delta \bar{W}_{\chi}
\end{pmatrix} .
\end{equation}
The two eigenvalues $\lambda_{\pm}$ corresponding to the coefficient matrix are given by
\begin{eqnarray}
\lambda_{\pm} &=& \frac12 \Big(1 - 6M^2 +  M/m \Big) \left(1 \pm
\sqrt{ 1 - \frac{16  \bar{m}_{\chi}^2  M^4}{ \kappa(\phi) (1 - 6M^2 +  M/m)^2 }}
\right)\\
&\simeq&
\Big(1 - 3M^2 \Big) \left(1 \pm
\sqrt{ 1 - \frac{0.8  m_{\chi}^2  M^6 e^{2 t}}{  (1-6M^2) (1 - 3M^2 )^2 }}
\right),
\end{eqnarray}
where we substituted $\bar{V}_0=0.1$ and $M/m=1.015$ as defined in Eq.~\eqref{boc} and used $\kappa(\phi) = (-\bar{V}/\bar{V}_0)^{M/m}$ to evaluate $\kappa(\phi)$ at the flat FRW stationary point.
%{\color{red} Need to recheck factors of $\Theta_0$ in that last expression.} I don't see any mistake here.
%
Note that the second term under the square root is positive  for $M\leq 0.2$ and $m_{\chi} = 300 \Theta_0{}^{-1}$ and becomes vanishingly small as  $t\to- \infty$.  

Although the analysis to this point might suggest that the flat FRW fixed point is stable, there remains the (rescaled) spatial gradient of $\chi$ to consider, which turns out to be the source of the instability.   More precisely, the same factor $\kappa(\phi)$  that stabilizes the time derivative of $\chi$ destabilizes the $\bar{S}_{\chi}{}^x\simeq0$ gradient contribution: Evaluating the evolution equation~\eqref{L-Schi-eq} of $\bar{S}_{\chi}{}^x$ for the flat FRW stationary point solution in Eq.~\eqref{FRW-sp}, 
 \begin{equation}
\label{sta-eq-Schi-sp}  
 {\partial _t} \delta \bar{S}_{\chi}{}^x \simeq \Big(1 - 2M^2 -   M/m \Big) \delta \bar{S}_{\chi}{}^x,
\end{equation}
with $M/m=1.015$, it becomes apparent that the gradient perturbation slowly grows,
\begin{equation} \label{eqSgrow}
\delta \bar{S}_{\chi}{}^x =\delta \bar{S}_{\chi}{}^x(t_{\rm FRW}) e^{- ( 0.015 + 2M^2 )\,t},
\end{equation} 
where $\delta \bar{S}_{\chi}{}^x(t_{\rm FRW})$ is the value of $\delta \bar{S}_{\chi}{}^x $ at the beginning of Stage 3, when the flat FRW stage begins.  This expression is valid at the linear level during Stage 3, the small entropic fluctuations shown in the last panel of Fig.~\ref{Figure5} arise as non-linear gradient effects on the evolution grow to become non-negligible. 

\vspace{.1in}
\noindent
{\it Stage 4:} 
$\bar{S}_{\chi}{}^x $ eventually grows large enough to deflect the evolution away from flat FRW and towards the stable Kasner-like fixed point with a time-invariant, non-zero $\delta \bar{S}_{\chi}{}^x$  where
\begin{alignat}{1}
\label{stat-point-n}
& {\cal N}^{-1} = \frac{6 -  3\,M/m  }{3 (M/m)^2 - 5 \,M/m + 2  + 8M^2}, \\ 
& \bar{W}_{\chi} = 0, \\
& \bar{S}_{\chi x}^2 = 2\,{\cal N}^{-2}\times\frac{ \Big(3 (M/m)^2  - 4\,M/m  + 8M^2\Big) \Big(1  - M/m  - 2M^2 \Big)}{(2-  M/m )^2}, 
 \\
 &\bar{n}_{ab} = 0 \;{\rm for\,all}\; a,b,\\
 & \bar{\Sigma}_{xx} = -\frac23 \times \frac{\bar{S}_{\chi x}^2}{{\cal N}^{-1}-3}, \\  
& \bar{\Sigma}_{yy} =  \bar{\Sigma}_{zz} = -\frac12 \times \bar{\Sigma}_{xx}  ,\\
&\bar{\Sigma}_{ab} = 0 \;{\rm for\,all}\; a\neq b,
\\
 & \bar{W}_{\phi} = 2\,m \times \left({\cal N}^{-1} -1 - \bar{\Sigma}_{xx} \right), \\
 \label{stat-point-Vbar}
& \bar{V}(\phi) = -M\times \left( \frac{m^{-1}}{2} \times \bar{S}_{\chi x}^2 + ({\cal N}^{-1}-3)\bar{W}_{\phi}  \right). 
\end{alignat}
As we show in Appendix~\ref{grad-stab}, the  Kasner-like stationary point corresponds to a state with a time-independent non-zero gradient, the true  attractor solution for combinations of $m_{\chi}, M$, and $m$ employed in our studies.

Eq.~\eqref{eqSgrow} is a key result:  It tells us that the characteristic duration of Stage 3 ({\it i.e.}, the period where the evolution remains close to the unstable flat FRW fixed point) is of ${\cal O}(\frac{1}{ 0.015 + 2M^2 })$, which can be more than one hundred $e$-foldings for values of $M$ that are not pushed too close to the Planck scale and for even small values of $m_{\chi}$, as shown in the example in Fig.~\ref{Figure4}.  The precise duration depends on the value of $\delta \bar{S}_{\chi}{}^x(t_{\rm FRW}) $, the magnitude of $\delta \bar{S}_{\chi}{}^x $ at the beginning of  Stage 3.  The  value depends on the initial conditions and the non-linear evolution that takes places in Stages~1 and~2, which can only be determined by evolving the system of equations using full numerical relativity. The important point, though, is that Stage 3, the flat FRW phase,  endures sufficiently long for values of $M$ and $m_{\chi}$ characteristic of    
bouncing cosmologies with a smooth (non-singular) bounce; the later instability is irrelevant because the bounce would occur while the spacetime is still close to the flat FRW state.

\section{Discussion}
\label{sec_conclusion}

Using the tools of numerical general relativity, we studied the cosmological evolution during slow contraction in models where the stress-energy is sourced by two kinetically-coupled scalar fields with an exponential non-linear $\sigma$ model-type interaction beginning from inhomogeneous and anisotropic initial conditions that deviate far from flat FRW spacetimes.  Our main finding was that slow contraction led to a robust and rapid convergence to a flat FRW geometry, similar to previous results obtained in Refs.~\cite{Cook:2020oaj,Ijjas:2020dws, Ijjas:2021gkf} for the case where slow contraction is sourced by a single canonical scalar field. 

However, our study revealed a subtle difference as well. Whereas the flat FRW solution is a stable attractor in the single-field case, it is not in the case of the two kinetically coupled fields considered here.  Instead, the evolution in the two-field case rapidly becomes ultralocal and approaches close to the flat FRW fixed point where it remains for a considerable period, but ultimately it is deflected away from that fixed point and towards a Kasner-like fixed point.  We investigated the instability analytically and showed that the characteristic time for the instability to develop enough to cause the deflection is ${\cal O}(100)$ or more $e$-folds of contraction of the inverse mean curvature $\Theta$.  

In addition, we showed that slightly decreasing the characteristic scale $M$ of the negative exponential potential $\bar{V} = -\bar{V}_0e^{-\phi/M}$ associated with the field $\phi$ that drives slow contraction as well as increasing the mass $\bar{m}_{\chi}$ of the light $\chi$ field which couples to $\phi$ through a non-linear $\sigma$ type kinetic interaction further increases the duration of the flat FRW period and delays the deflection to the Kasner-like fixed point.  

The result is reminiscent of the single-field case discussed in Ref.~\cite{Ijjas:2021wml} where, for some initial data sets, we found that the system first evolved close to a Kasner-like fixed point  before it was deflected to the stable flat FRW attractor solution. There are two key differences between the two cases, though: 
\begin{itemize}
\item[-] in the single field case, the system approaches close to the (unstable) Kasner-like fixed point for only a {\it very small subset of initial conditions} and it remains there for only a {\it  short period} before being deflected to the stable attractor fixed point (the flat FRW state), 
\item[-] in the the kinetically-coupled two-field case, the system approaches close to the (unstable) flat FRW fixed point  for a {\it very wide range of initial conditions} and  remains there for a {\it long period} before being deflected to the stable attractor Kasner-like fixed point. 
\end{itemize}

This result is surprising and was not anticipated in previous studies based on the conventional perturbative techniques commonly applied in cosmology.
Hence, this study is a fine demonstration of the power of non-perturbative, numerical relativity to reveal novel cosmological dynamics that one could not anticipate through conventional perturbative methods but that are critically important to understand  in developing cosmological models.  

In the present example, the difference between the single-field and kinetically-coupled two-field case turns out not to be observationally relevant in bouncing cosmologies because the end of slow contraction and the transition to the hot expanding phase begins around $100$ $e$-foldings of contraction of $\Theta$ -- before the flat FRW state destabilizes.  This is significant because the kinetically-coupled two-field models studied here are examples that can generate a nearly scale-invariant spectrum of super-Hubble density fluctuations fully consistent with cosmic microwave background observations.

%%%%%%

\section*{Acknowledgments}
The work of A.I. is supported by the Lise Meitner Excellence Program of the Max Planck Society and by the Simons Foundation grant number 663083.
F.P. acknowledges support from NSF grant PHY-1912171, the Simons Foundation, and the Canadian Institute For Advanced Research (CIFAR).  P.J.S. is supported in part by the DOE grant number DEFG02-91ER40671 and by the Simons Foundation grant number 654561.  D.G. is supported by the NSF grants PHY-1806219 and PHY-2102914.

\appendix

\section{Tetrad frame transformation rules}
\label{app:frame}

In this Appendix, we consider the relationship between geometric variables as represented in the hypersurface-orthogonal tetrad frame (used in our tetrad codes to study smoothing of initial conditions) and the the co-moving tetrad frame. To this end, we will first derive the transformation rules for the tetrad vector components, the Ricci rotation coefficients and the (effective) `fluid' variables describing the stress-energy under the Lorentz boost $\{\Lambda_{\alpha}{}^{\beta}\}$ that connects two arbitrary tetrad frames. Then, we apply the rules to the case of scalar fields to transform geometric and `fluid' variables from the hypersurface-orthogonal tetrad frame to the co-moving tetrad frame.

We are interested in tetrad frame transformations under Lorentz boosts $\{\Lambda_{\alpha}{}^{\beta}\}$ that transform a time-like tetrad ${\bf e_0}$ to another ${\bf \tilde{e}}_0$. The boost is defined through the rapidity $\beta$ between the two 4-vectors ${\bf e_0}$ and ${\bf \tilde{e}}_0$ 
and through the projection $\{w_a\}$ of ${\bf \tilde{e}}_0$ into the 3-surfaces spanned by the spatial triad ${\bf e_a}$ ($a=1,2,3$):
 \begin{equation}
 \label{boost-def}
\Lambda_0{}^0  %= - u_{\alpha}\,e_0{}^{\alpha} 
\equiv {\rm cosh}\beta , \quad 
\Lambda_a{}^0 \equiv {\rm sinh} \beta\, w_a , \quad 
\Lambda_a{}^b  = w_aw^b({\rm cosh}\beta -1) + \delta_a{}^b\,,
\end{equation}
where $w_aw^a=1$. 
Note that $\Gamma \equiv  {\rm cosh}\beta = 1/\sqrt{1-{\bf v}^2}$ is the Lorentz factor %(or peculiar velocity of the `fluid') 
with $v_a \equiv {\rm tanh} \beta\, w_a $. %is the fluid's 3-velocity. 
In particular, 
\begin{equation}
 \label{boost-def-vgamma}
\Lambda_0{}^0  = \Gamma, \quad 
\Lambda_a{}^0 = \Gamma v_a , \quad 
\Lambda_a{}^b  = \frac{\Gamma^2}{\Gamma+1}v_av^b + \delta_a{}^b\,.
\end{equation}

\vspace*{0.1in}
\noindent
{\it Vierbein.}
The tetrad frame vectors $\{e_{\alpha}\}$ and $\{\tilde{e}_{\alpha}\}$ (with $\alpha=0,...,3$) are related as follows:
\begin{eqnarray}
\tilde{e}_0 &=& \Lambda_0{}^{\beta} {\bf e}_{\beta} =  \Gamma\Big({\bf e}_0 + v^b{\bf e}_b\Big)
\,,\\
\tilde{e}_a &=& \Lambda_a{}^{\beta} e_{\beta}
=  \Gamma v_a {\bf e}_0 + \Big(\delta_a{}^b + \frac{\Gamma^2}{\Gamma + 1}v_a v^b\Big){\bf e}_b
\,.
\end{eqnarray}
%where $\Lambda_{\alpha}{}^{\beta}$ is a Lorentz matrix with $\Lambda_0{}^0>0, {\rm Det}(\Lambda)=1$ and 
%\begin{equation}
%\eta_{\alpha\beta} = \Lambda_{\alpha}{}^{\gamma}\Lambda_{\delta}{}^{\beta} \eta_{\gamma\delta}\,.
%\end{equation}

\vspace*{0.1in}
\noindent
{\it Ricci Rotation Coefficients.}
The Ricci Rotation Coefficients $\gamma_{\alpha\beta\lambda}$ which are defined as
\begin{equation}
\gamma_{\alpha\beta\lambda} \equiv e_{\alpha} \nabla_{\lambda} e_{\beta},
\end{equation}
with $\nabla_{\lambda} \equiv e_{\lambda}{}^{\mu}\nabla_{\mu}$ transform as follows:\begin{eqnarray}
\tilde{\gamma}_{\alpha\beta\lambda} &=& \tilde{e}_{\alpha} \tilde{\nabla}_{\lambda} \tilde{e}_{\beta}
= \tilde{e}_{\alpha} \tilde{e}_{\lambda}{}^{\mu} \tilde{\nabla}_{\mu} \tilde{e}_{\beta}
= \Lambda_{\alpha}{}^{\delta} e_{\delta} \Lambda_{\lambda}{}^{\zeta} e_{\zeta}{}^{\mu} \nabla_{\mu} \left( \Lambda_{\beta}{}^{\epsilon} e_{\epsilon} \right)\\
&=& \Lambda_{\alpha}{}^{\delta} \Lambda_{\beta}{}^{\epsilon} \Lambda_{\lambda}{}^{\zeta} \gamma_{\delta\epsilon\zeta} 
- \eta_{\delta\epsilon}  \Lambda_{\beta}{}^{\epsilon} \Lambda_{\lambda}{}^{\zeta}\nabla_{\zeta} \Lambda_{\alpha}{}^{\delta}\,.
\nonumber
\end{eqnarray}

\vspace*{0.1in}
\noindent
{\it Components of the Stress-Energy Tensor.}
Characterizing an arbitrary stress-energy tensor through effective `fluid' variables,
\begin{eqnarray}
\varrho &\equiv& e_0{}^\mu e_0{}^\nu T_{\mu\nu} 
 ,\\
j_a &\equiv& - e_0{}^\mu e_a{}^\nu T_{\mu\nu}   
,\\
s_{ab} &\equiv& \pi_{ab} + p\delta_{ab} \equiv e_a{}^\mu e_b{}^\nu T_{\mu\nu}  
 ,\\
p &\equiv& {\textstyle \frac13} s_a{}^a 
 \,,
\end{eqnarray}
where $\varrho$ is the energy density, $j_a$ the three-momentum flux, $s_{ab}$ the spatial stress tensor and $p$ denotes the pressure and $\pi_{ab}$ denotes the trace-free anisotropic stress (with $\pi_{ab}\equiv s_{(ab)}-{\textstyle \frac13} s_a{}^a \delta_{ab}$),
under a Lorentz boost as defined in Eq.~\eqref{boost-def}, the `fluid' variables transform as follows:
\begin{eqnarray}
\tilde{T}_{\alpha\beta} &=& \tilde{e}_{\alpha}{}^{\mu} \tilde{e}_{\beta}{}^{\nu} T_{\mu\nu}
=  \Lambda_{\alpha}{}^{\gamma} \Lambda_{\beta}{}^{\delta} e_{\gamma}{}^{\mu}e_{\delta}{}^{\nu}T_{\mu\nu}\\
&=& \Lambda_{\alpha}{}^0 \Lambda_{\beta}{}^0 \rho
- \Lambda_{\alpha}{}^0 \Lambda_{\beta}{}^a j_a
-\Lambda_{\alpha}{}^b \Lambda_{\beta}{}^0 j_b
+\Lambda_{\alpha}{}^a \Lambda_{\beta}{}^b s_{ab}
\nonumber,
\end{eqnarray}
{\it i.e.},
\begin{eqnarray}
\tilde{ \varrho} &=& 
%\tilde{e}_0{}^\mu \tilde{e}_0{}^\nu T_{\mu\nu} 
%= \Lambda_0{}^{\gamma} \Lambda_0{}^{\delta} T_{\gamma\delta}= 
\Gamma^2 \varrho - 2\Gamma^2j_av^a + \Gamma^2v^av^b s_{ab}
 %= {\textstyle \frac12} D_0 \phi D_0\phi  + {\textstyle \frac12} D_a \phi D^a\phi + V(\phi) 
 ,\\
\tilde{ j}_a &=& 
%- \tilde{e}_0{}^\mu \tilde{e}_a{}^\nu T_{\mu\nu}   
%= - \Lambda_0{}^{\gamma} \Lambda_a{}^{\delta} T_{\gamma\delta} = 
-\Gamma^2 \varrho v_a + \Gamma j_a + \Gamma^2 \left( 1 + \frac{\Gamma}{\Gamma+1}\right) j_bv^b v_a - \Gamma \left( \delta_a{}^c + \frac{\Gamma^2}{\Gamma+1}v_av^c\right) v^b s_{bc}
%=- D_0\phi D_a\phi
,\\
 \tilde{s}_{ab} &=& 
 %\tilde{e}_a{}^\mu \tilde{e}_b{}^\nu T_{\mu\nu}  = \Lambda_a{}^{\gamma} \Lambda_b{}^{\delta} T_{\gamma\delta}
 s_{ab} +  \Gamma^2 \varrho v_av_b - \Gamma \Big(v_aj_b + v_bj_a\Big) - 2\frac{\Gamma^3}{\Gamma+1} v^cj_cv_av_b   
 + \frac{\Gamma^4}{(\Gamma+1)^2}v^cv^ds_{cd}v_av_b
 \\
  &+&  \frac{\Gamma^2}{\Gamma+1}\Big(v_bv^c s_{ac} + v_av^cs_{cb}\Big)
 \nonumber
 %= D_a \phi D_b\phi  + \left({\textstyle \frac12} D_0 \phi D_0\phi  - D_c \phi D^c\phi - V(\phi) \right)\delta_{ab}
 \,.
\end{eqnarray}

%\subsection*{From hypersurface-orthogonal to co-moving frame}

For the numerical relativity codes that we use to study the robustness and rapidity of slow contraction, we fixed the tetrad frame gauge to be hypersurface-orthogonal and Fermi propagated, as described in Sec.~\ref{sec:gauge}. In particular, the timelike vierbein ${\bf e_0}$ is normal to spacelike hypersurfaces. In general, $j_a\neq0$ in this frame, which means  that, typically, ${\bf e_0}$ and the effective fluid's 4-velocity (which is the timelike congruence ${\bf \tilde{e}}_0$ of the co-moving tetrad) do not coincide.

For example, in the case of a single scalar field, where 
\begin{equation}
T_{\mu\nu} = \nabla_{\mu}\phi\nabla_{\nu}\phi - \Big({\textstyle \frac12}\nabla_{\mu}\phi\nabla^{\mu}\phi + V(\phi)\Big)g_{\mu\nu},
\end{equation}
the `fluid' variables $\varrho, j_a, s_{ab}$, and $p$ in the hypersurface-orthogonal tetrad frame take the following form:
 \begin{eqnarray}
\varrho &=& 
 {\textstyle \frac12} D_0 \phi D_0\phi  + {\textstyle \frac12} D_a \phi D^a\phi + V(\phi) 
 ,\\
j_a &=& - D_0\phi D_a\phi
,\\
 s_{ab} &=& 
 D_a \phi D_b\phi  + \Big({\textstyle \frac12} D_0 \phi D_0\phi  - {\textstyle \frac12} D_c \phi D^c\phi - V(\phi) \Big)\delta_{ab},\\
p &=& {\textstyle \frac12} D_0 \phi D_0\phi  - {\textstyle \frac16} D_c \phi D^c\phi - V(\phi)
 \,.
\end{eqnarray}
Here, $D_0$ denotes the Lie derivative along $e_0$ and $D_a$ is the directional derivative along $e_a$. Manifestly, ${\bf e}_0 \neq - D_a\phi/D_0\phi$ ad hence the hypersurface-orthogonal and co-moving tetrad frames do not coincide in general.
The Lorentz boost, $v_a = - D_a\phi/D_0\phi$,  transforms the tetrad to the co-moving frame, where the stress-energy takes the form of a perfect fluid:
\begin{equation}
\tilde{\varrho} = {\textstyle \frac12} D_0 \phi D_0\phi  + {\textstyle \frac12} D_c \phi D^c\phi - V(\phi) , \quad \tilde{p} = {\textstyle \frac12} D_0 \phi D_0\phi  - {\textstyle \frac12} D_c \phi D^c\phi - V(\phi) ,\end{equation}
\begin{equation}
 j_a = \pi_{ab}=0.
\end{equation}
Note, though, that the hypersurface-orthogonal tetrad frame converges to the co-moving frame if  $v_a = (D_a \phi/\dot{\phi}) \to 0$, as is the case in our simulations that involve a single scalar which is minimally-coupled to gravity and has a negative potential. 
By contrast, our initial data as specified above in Sec.~\ref{sec:ID} is {\it not} represented in the co-moving frame. This does not affect the conclusion that the spacetime converges to a homogeneous, isotropic and spatially flat FRW universe.  However, it does mean that the initial conditions are not  as a Eulerian co-moving observer would measure them to be. 
%Significantly, this also means that, in cases where the evolution does not become ultralocal ({\it e.g.} in the case of `bad' inflation),  there is not convergence to a comoving frame and the values of  variables expressed in the  code's hypersurface-orthogonal tetrad frame are never the physically relevant ones.

More generally, the Lorentz boost $v_a = j_a/(\varrho + p_c)$ with $p_c$ being the co-moving pressure transforms an arbitrary tetrad frame to the co-moving frame. If, in addition, $s_{ab}=j_aj_b + p_c\delta_{ab}$, it is straightforward to show that the `fluid' takes the perfect fluid form in the co-moving tetrad frame with 
\begin{equation}
\tilde{\varrho} = \frac{1}{\Gamma^2}\big(\varrho+p_c\big) - p_c, \quad \tilde{s}_{ab} = p_c\delta_{ab},\quad j_a = \pi_{ab}=0.
\end{equation}

In the case of two kinetically-interacting scalars like those considered in this paper with 
\begin{equation}
T_{\mu\nu} = \nabla_{\mu}\phi\nabla_{\nu}\phi + \kappa(\phi)\nabla_{\mu}\chi\nabla_{\nu}\chi - \Big({\textstyle \frac12}\nabla_{\mu}\phi\nabla^{\mu}\phi + {\textstyle \frac12}\kappa(\phi)\nabla_{\mu}\chi\nabla^{\mu}\chi + V(\phi) + U(\chi)\Big)g_{\mu\nu},
\end{equation}
the `fluid' variables $\varrho, j_a, s_{ab}$, and $p$ in the hypersurface-orthogonal tetrad frame take the following form:
 \begin{eqnarray}
\varrho &=& 
 {\textstyle \frac12} \Big(D_0 \phi D_0\phi  + D_a \phi D^a\phi \Big) + V(\phi)
 + {\textstyle \frac12}\kappa(\phi)\Big( D_0 \chi D_0\chi  + D_a \chi D^a\chi\Big) + U(\chi) 
 ,\\
 \label{ja-twofield}
j_a &=& - D_0\phi D_a\phi - \kappa(\phi) D_0\chi D_a\chi
,\\
 s_{ab} &=& 
 D_a \phi D_b\phi  + \kappa(\phi) D_a \chi D_b\chi  \\
& +&  \Big({\textstyle \frac12} \big(D_0 \phi D_0\phi  -  D_c \phi D^c\phi \big) - V(\phi) + {\textstyle \frac12} \kappa(\phi) \big(D_0 \chi D_0\chi  -  D_c \chi D^c\chi\big) - U(\chi) \Big)\delta_{ab},\nonumber\\
p &=& {\textstyle \frac12} \Big(D_0 \phi D_0\phi  - {\textstyle \frac13} D_c \phi D^c\phi\Big) - V(\phi)
+ {\textstyle \frac12} \kappa(\phi)\Big( D_0 \chi D_0\chi  - {\textstyle \frac13} D_c \chi D^c\chi\Big) - U(\chi)
 \,.
\end{eqnarray}
Again, it is immediately apparent from Eq.~\eqref{ja-twofield} that the hypersurface orthogonal and the co-moving frames do not coincide.

The Lorentz boost defined through
\begin{equation}
\label{va-twofield}
v_a = - \frac{D_0\phi D_a\phi + \kappa(\phi) D_0\chi D_a\chi}{D_0 \phi D_0\phi +\kappa(\phi) D_0 \chi D_0\chi } = \frac{j_a}{\rho + p_c}
\end{equation} 
transforms the hypersurface-orthogonal tetrad to the co-moving frame. Here, the co-moving pressure is defined as
\begin{equation}
p_c \equiv {\textstyle \frac12} \Big(D_0 \phi D_0\phi  -  D_c \phi D^c\phi\Big) - V(\phi)
+ {\textstyle \frac12} \kappa(\phi)\Big( D_0 \chi D_0\chi  -  D_c \chi D^c\chi\Big) - U(\chi),
\end{equation}
and the co-moving `fluid' variables take the following form:
\begin{eqnarray}
\tilde{ \varrho} &=& 
\rho + (1-\Gamma^2) (\rho + p_c)  + \Gamma^2  \big( D^a\phi D^b\phi + \kappa(\phi)D^a\chi D^b\chi \big)  \frac{j_aj_b}{(\rho + p_c)^2} 
 ,\\
\tilde{ j}_a &=& 
\big(\Gamma^2 - 1 \big) j_a 
- \Gamma  \big( D_a\phi D^b\phi + \kappa(\phi)D_a\chi D^b\chi \big)  \frac{j_b}{\rho + p_c} 
\\
&-& \frac{\Gamma^3}{\Gamma+1}\big( D^b\phi D^c\phi + \kappa(\phi)D^b\chi D^c\chi \big)  \frac{ j_bj_c j_a}{(\rho + p_c )^3}  ,
\nonumber\\
 \tilde{s}_{ab} &=& 
 p_c\delta_{ab} + D_a\phi D_b\phi + \kappa(\phi)D_a\chi D_b\chi 
 -\Gamma^2 \frac{j_a j_b}{\rho + p_c}
\\
& +& \frac{\Gamma^2}{\Gamma+1} \left( \big( D_b\phi D^c\phi + \kappa(\phi)D_b\chi D^c\chi \big)  \frac{ j_c j_a}{(\rho + p_c )^2}
+  \big( D_a\phi D^c\phi + \kappa(\phi)D_a\chi D^c\chi \big)  \frac{ j_c j_b}{(\rho + p_c )^2}  \right)
\nonumber\\
& +& \frac{\Gamma^4}{(\Gamma+1)^2} \big( D^c\phi D^d\phi + \kappa(\phi)D^c\chi D^d\chi \big)  \frac{ j_cj_d j_aj_b}{(\rho + p_c )^4}\nonumber
 \,.
\end{eqnarray}
In general, interacting scalar fields act as `imperfect fluids' with non-zero momentum flux and non-diagonal spatial stress tensor.

In our studies, we found that all solutions evolve towards stationary points with 
\begin{itemize}
\item[-] a homogeneous $\phi$ profile, {\it i.e.}, $D_a\phi \equiv 0$ for all $a\in\{1,2,3\}$; and
\item[-] a non-dynamical $\chi$ profile, {\it i.e.},  $D_0\chi \equiv 0$. 
\end{itemize}
Accordingly, for all solutions, $v_a$ as defined in Eq.~\eqref{va-twofield} evolves to zero. This means, the hypersurface orthogonal frame evolves towards the co-moving frame such that all geometric and `fluid' variables in the hypersurface-orthogonal frame faithfully represent the Eulerian observer's measurements. In particular,
\begin{eqnarray}
\rho &\to& \tilde{ \varrho} = {\textstyle \frac12} D_0 \phi D_0\phi + V(\phi)
 + {\textstyle \frac12}\kappa(\phi)D_a \chi D^a\chi + U(\chi) 
 ,\\
j_a &\to& \tilde{ j}_a = 0 
,\\
 s_{ab} &\to& \tilde{s}_{ab} = 
 p_c\delta_{ab} + \kappa(\phi)D_a\chi D_b\chi \,.
 \end{eqnarray}

\section{Dynamical stability of the fixed point solution with $\bar{S}_{\chi}{}^x\neq0, \partial_t \bar{S}_{\chi}{}^x=0$}
\label{grad-stab}

To analyze the stability of the new stationary solution, we linearize the Einstein-scalar system~(\ref{E-eq}-\ref{Wchi-eq}) around the fixed point solution given in Eq.~\eqref{stat-point-n}:
\begin{alignat}{2}
%E_alpha^i
\label{E-eq-L2}
&\partial _t \delta \bar{E}_a{}^i &{}={}& \Big(1- {\cal N} \big( 1 + \bar{\Sigma}_{aa}\big) \Big) \delta \bar{E}_a{}^i 
,\\
%A_alpha
&{\partial _t} \delta\bar{A}_b &{}={}& \Big(1- {\cal N}\big( 1 + \bar{\Sigma}_{bb}\big) \Big) \delta\bar{A}_b 
,\\
%N^alphabeta
\label{dn-ab-sta}
& \partial _t \delta \bar{n}_{ab} &{}={}& \Big(1 - {\cal N} \big( 1 -\bar{\Sigma}_{aa} - \bar{\Sigma}_{b b}\big)\Big) \delta \bar{n}_{ab} 
,\\
%delta Sigma_xx
& \partial _t \delta \bar{\Sigma}_{xx} &{} ={} & \Big(1 - 3 {\cal N} \Big)\delta \bar{\Sigma}_{xx} - \left( 3  \bar{\Sigma}_{xx} - \frac23  (\bar{S}_{\chi}{}^x)^2\right) \delta {\cal N} 
+ \frac43 {\cal N} \bar{S}_{\chi}{}^x\delta \bar{S}_{\chi}{}^x
,\\
%delta Sigma_yy
& \partial _t \delta \bar{\Sigma}_{yy} &{} ={} & \Big(1 - 3 {\cal N} \Big)\delta \bar{\Sigma}_{yy} - \left( 3  \bar{\Sigma}_{yy} + \frac13  (\bar{S}_{\chi}{}^x)^2\right) \delta {\cal N}
-\frac23 {\cal N} \bar{S}_{\chi}{}^x\delta \bar{S}_{\chi}{}^x
,\\
%delta Sigma_xy
\label{dsigma-xy-sta}
& \partial _t \delta \bar{\Sigma}_{xy} &{} ={} & \Big(1 - 3 {\cal N} \Big)\delta \bar{\Sigma}_{xy} 
,\\
%W_phi
&{\partial _t} \delta \bar{W}_{\phi} &{}={}& \Big(1 - 3 {\cal N} \Big) \delta \bar{W}_{\phi}    
+ {\cal N} \Big( M^{-1} \delta \bar{V} + m^{-1}  \bar{S}_{\chi}{}^x\delta  \bar{S}_{\chi}{}^x \Big)
\\
&& -&  \left( 3 \bar{W}_{\phi} -M^{-1} \bar{V} - \frac12 m^{-1}  (\bar{S}_{\chi}{}^x)^2 \right ) \delta {\cal N}
,\nonumber\\
%chi-eq
\label{L-chi-eq2}
&\partial _t \delta \chi &{}={}& {\cal N} \,\frac{ \delta \bar{W}_{\chi}}{\sqrt{\kappa(\phi)}}
,\\
%Wchi-eq
\label{L-Wchi-eq2}
 &   {\partial _t} \delta \bar{W}_{\chi} &{}={}& \left(1- {\cal N}\left(3 - \frac12 m^{-1}  \bar{W}_{\phi}\right)  \right) \delta \bar{W}_{\chi} 
- {\cal N} \frac{ \bar{U}_{,\chi\chi}}{\sqrt{\kappa}}\delta\chi
,\\
%S_chi
& {\partial _t} \delta \bar{S}_{\chi}{}^x &{}={}& 
\Big(1 - {\cal N} \left(1 +  \frac12 m^{-1} \bar{W}_{\phi} + \bar{\Sigma}_{xx} \right) \Big) \delta \bar{S}_{\chi}{}^x 
 - {\cal N} \bar{S}_{\chi}{}^x \left(\frac12 m^{-1}\delta\bar{W}_{\phi} + \delta\bar{\Sigma}_{xx} \right)  \\
 & &{}-{}&  \Big(1 +  \frac12 m^{-1} \bar{W}_{\phi} + \bar{\Sigma}_{xx} \Big)  \bar{S}_{\chi}{}^x \delta{\cal N} 
\nonumber
\end{alignat}
with $\delta {\cal N}$ and $\delta {\bar V}(\phi)$ being given through the linearized lapse equation and Hamiltonian constraint:
\begin{alignat}{1}
 &\delta {\bar V}(\phi)  = - \bar{\Sigma}_{xx}  \big(\delta \bar{\Sigma}_{xx} -\delta \bar{\Sigma}_{yy} \big)   -  \bar{W}_{\phi}\delta \bar{W}_{\phi}
- \bar{S}_{\chi}{}^x\delta \bar{S}_{\chi}{}^x
,\\
\label{Neqn-rs-limit-p}
& \delta {\cal N}  = 
 %- \frac13 {\cal N}^2 \Big(2\bar{\Sigma}_{xx}  \delta \bar{\Sigma}_{xx} + 4\bar{\Sigma}_{yy}  \delta \bar{\Sigma}_{yy} + 2\bar{W}_{\phi}  \delta \bar{W}_{\phi}   - \delta {\bar V}(\phi)\Big) \\&= 
 -  {\cal N}^2 \left(
\bar{\Sigma}_{xx}  \big( \delta \bar{\Sigma}_{xx} - \delta \bar{\Sigma}_{yy} \big)
+ \bar{W}_{\phi}  \delta \bar{W}_{\phi}  + \frac13\bar{S}_{\chi}{}^x\delta \bar{S}_{\chi}{}^x\right).%\nonumber
\end{alignat}

For $M/m=1.015$ and $M\leq0.2$, the stationary point solution given in Eq.~\eqref{stat-point-n} can be approximated as follows:
\begin{alignat}{1}
\label{stat-point-n-approx}
& {\cal N} \approx \frac83 \times M^2 , \\ 
& \bar{W}_{\chi} = 0, \\
& \bar{S}_{\chi}{}^x \approx \frac{3}{4}\times M^{-1} 
,  \\
 &\bar{n}_{ab} = 0 \;{\rm for\,all}\; a,b,
 \\
 \label{stat-point-sigxx-approx}
 & \bar{\Sigma}_{xx} \approx -\big( 1 + 0.005 M^{-2}\big), \\  
  \label{stat-point-sigyy-approx}
& \bar{\Sigma}_{yy} =  \bar{\Sigma}_{zz} \approx \frac12 \times \big( 1 + 0.005 M^{-2}\big) ,\\
&\bar{\Sigma}_{ab} = 0 \;{\rm for\,all}\; a\neq b,
\\
 & \bar{W}_{\phi} \approx  \frac34  \times M^{-1} , \\
 \label{stat-point-Vbar}
& \bar{V}(\phi) \approx -\frac{9}{16} \left( M^{-2} - 4 \right). 
\end{alignat}

It is immediately apparent from Eqs.~(\ref{E-eq-L2}-\ref{dn-ab-sta}, \ref{dsigma-xy-sta}) that $\delta  \bar{E}_a{}^i, \delta \bar{A}_b, \delta \bar{n}_{ab}$ and $\delta \bar{\Sigma}_{xy}$ decouple and evolve as follows: 
\begin{alignat}{1}
\delta \bar{E}_a{}^i & \propto e^{\left(1 - {\cal N} \big( 1 -\bar{\Sigma}_{aa}\big) \right) t }
,\\
\delta \bar{A}_b & \propto e^{\left(1 - {\cal N} \big( 1 -\bar{\Sigma}_{bb}\big) \right) t }
,\\
\delta \bar{n}_{ab} &\propto e^{\left(1 - {\cal N} \big( 1 -\bar{\Sigma}_{aa} - \bar{\Sigma}_{b b}\big)\right)\, t }
,\\
\delta \bar{\Sigma}_{xy} &\propto e^{ \big(1 - 3 {\cal N} \big)\,t }.
\end{alignat}
Substituting Eqs.~(\ref{stat-point-n-approx}, \ref{stat-point-sigxx-approx}-\ref{stat-point-sigyy-approx}) for ${\cal N}, \bar{\Sigma}_{xx}$ and $\bar{\Sigma}_{yy}$, respectively, it is straightforward to verify that all the coefficients of $t$ in the exponents are all positive, such that corresponding perturbations decay as $t \rightarrow -\infty$.

Similarly, the linearized equations~(\ref{L-chi-eq2}-\ref{L-Wchi-eq2}) for $\delta \chi$ and $\delta\bar{W}_{\chi}$ form a closed system,
\begin{equation}
 \partial_t \begin{pmatrix}
\delta \chi\\[1em]
\delta \bar{W}_{\chi}
\end{pmatrix} =
\begin{pmatrix}
0 & & (8/3)M^2/\sqrt{\kappa(\phi)}\\[1em]
-(8/3) M^2\bar{m}_{\chi}^2/\sqrt{\kappa(\phi)} & {\quad} & 1 - 8M^2 +  M/m
\end{pmatrix}
\begin{pmatrix}
\delta \chi\\[1em]
\delta \bar{W}_{\chi}
\end{pmatrix},
\end{equation}
which is being continuously damped as $|t|$ grows since both eigenvalues 
\begin{equation}
\lambda_{\pm} \approx
\Big(1 - 4M^2 \Big) \left(1 \pm
\sqrt{ 1 - \frac{ m_{\chi}^2 M^6 e^{2 t}}{  (1 - 4M^2 )^3 }}
\right)
\end{equation}
corresponding to the coefficient matrix
are both positive definite for $M\leq 0.2$. For example, for $M=0.1, m_{\chi}=300 \Theta_0^{-1}$ and $t=-2$, $\lambda_{+}\approx 1.9$ and $\lambda_{-}\approx 0.01$.  Note that to evaluate $\kappa(\phi)$ at the Kasner-like stationary point, we used $\kappa(\phi) = (-\bar{V}/\bar{V}_0)^{M/m}$ and substituted $\bar{V}_0=0.1$ as defined in Eq.~\eqref{boc}. 

The remaining four linearized variables $\delta \bar{\Sigma}_{xx}, \delta \bar{\Sigma}_{yy}, \delta \bar{W}_{\phi}$ and $\delta \bar{S}_{\chi}{}^x$ are determined by the closed system
\begin{alignat}{2}
%delta Sigma_xx
\label{lin-sig22-eq}
& \partial _t \delta \bar{\Sigma}_{xx} &{} \approx{} & 
\Big(1 - 5M^2 \Big)\delta \bar{\Sigma}_{xx} 
+ \frac{4}{3}\Big(M^2 + 0.005 \Big) \delta \bar{\Sigma}_{yy} 
-2M \delta \bar{W}_{\phi}
+ 2M \delta \bar{S}_{\chi}{}^x
,\\
%delta Sigma_yy
& \partial _t \delta \bar{\Sigma}_{yy} &{} \approx {} & 
- \frac43 \Big( M^2 + 0.005  \Big)\delta \bar{\Sigma}_{xx}
+ \Big(1 - 7M^2 \Big)\delta \bar{\Sigma}_{yy} 
+ M \delta \bar{W}_{\phi} - M \delta \bar{S}_{\chi}{}^x
,\\
%W_phi
&{\partial _t} \delta \bar{W}_{\phi} &{}\approx{}&   
 \frac23 M   \big( 1 + 0.005 M^{-2}\big)  \big(\delta \bar{\Sigma}_{xx} -\delta \bar{\Sigma}_{yy} \big) 
+ \Big(\frac12 - 8M^2 \Big) \delta \bar{W}_{\phi}  
+  \frac12 \delta \bar{S}_{\chi}{}^x
,\\
%S_chi
\label{lin-Schi-eq}
& {\partial _t} \delta \bar{S}_{\chi}{}^x &{}\approx{}&  - M \Big( 4 + 0.01M^{-2}\Big) \delta\bar{\Sigma}_{xx}   
 +  M   \big( 1 + 0.005M^{-2}  \big)\delta \bar{\Sigma}_{yy}
 + \frac12  \delta\bar{W}_{\phi} + \frac12 \delta \bar{S}_{\chi}{}^x 
.
\end{alignat}

It is straightforward to verify using, {\it e.g.}, a symbol mathematical computer program, that for $M\leq0.2$, all four eigenvalues $\lambda_{1,2,3,4}$ of the coefficient matrix corresponding to the system of ordinary differential equations~(\ref{lin-sig22-eq}-\ref{lin-Schi-eq}) have a positive real part. For example, for $M=0.1$, $\lambda_1 \approx 0.94, \lambda_2 \approx 0.91 , \lambda_3\approx 0.80$, and $\lambda_4\approx 0.13$.  Since our time coordinate $t$ is negative, running towards $-\infty$, a positive real part for all eigenvalues means all perturbations decay as $|t|$ grows and thus the Kasner-like stationary point solution with $\bar{S}_{\chi}{}^x \approx (3/4)M^{-1}$ is a stable attractor.

\bibliographystyle{plain}
\bibliography{twofield}

\end{document}